%
%
%

\def\oddpage{{\tt preliminary draft \hfil \jobname \hfil \today}}
\def\evenpage{{\tt \today \hfil\jobname \hfil preliminary draft}}
\def\titlepage{{\tt preliminary draft \hfill \jobname}}
\def\draft{\baselineskip = 16pt plus 2pt minus 1pt
   \hsize = 17.0 truecm \vsize = 24.7 truecm
   \hoffset=-4 truemm   
    \overfullrule = 5pt
    \headline{\ifnum\count0>1\ifodd\count0\oddpage\else%
    \evenpage\fi\else\titlepage\fi}
}
\overfullrule=0pt             
\hoffset=-3 true mm           
\voffset=9.5 true mm

%
%
%
%

\newif\iflabels

\newif\ifreference


\newcount\secnum    \global\secnum=0
\newcount\subsecnum \global\subsecnum=0
\newcount\eqnum     \global\eqnum=0
\newcount\citenum   \global\citenum=0
\newcount\fignum    \global\fignum=0
\newcount\tablenum  \global\tablenum=0
\newcount\remarknum    \global\remarknum=0

\newif\ifndouble
\def\doublenumbers{\ndoubletrue\gdef\therunningsection{\the\secnum}}

\def\ifundefined#1{\expandafter\ifx\csname#1\endcsname\relax}
\def\strip#1>{}


\def\cref#1{\ifundefined{@c@#1}\immediate\write16{ --> \string\cref{#1}
    not defined !!!}
    \expandafter\xdef\csname@c@#1\endcsname{??}\fi\csname@c@#1\endcsname}

\def\eqref#1{\ifundefined{@eq@#1}\immediate\write16{ --> \string\eqref{#1}
    not defined !!!}
    \expandafter\xdef\csname@eq@#1\endcsname{??}\fi\csname@eq@#1\endcsname}

\def\sref#1{\ifundefined{@s@#1}\immediate\write16{ --> \string\sref{#1}
    not defined !!!}
    \expandafter\xdef\csname@s@#1\endcsname{??}\fi\csname@s@#1\endcsname}

\def\figref#1{\ifundefined{@f@#1}\immediate\write16{ -->
    \string\figref{#1} not defined !!!}
    \expandafter\xdef\csname@f@#1\endcsname{??}\fi\csname@f@#1\endcsname}

\def\Fig#1{Fig.~\figref{#1}}
\def\Figure#1{Figure~\figref{#1}}

\def\tableref#1{\ifundefined{@f@#1}\immediate\write16{ -->
    \string\tableref{#1} not defined !!!}
    \expandafter\xdef\csname@f@#1\endcsname{??}\fi\csname@f@#1\endcsname}

\def\Tab#1{Table~\tableref{#1}}

\newdimen\beforesecskip  \beforesecskip=\baselineskip
\newdimen\aftersecskip   \aftersecskip=0pt

\font\sectionfont = cmbx10 at 11 pt

\def\Section#1#2{            
    \global\advance\secnum by 1\ifndouble\global\eqnum=0\fi
    \global\subsecnum=0
    \xdef\therunningsection{\the\secnum}
    \def\usenthrow{1}\ifundefined{@s@#1}\def\usenthrow{2}\fi
    \expandafter\ifx\csname@s@#1\endcsname\therunningsection\def\usenthrow{2}\fi
    \ifodd\usenthrow\immediate\write16
      { --> Possible reference error in \string\sref{#1} }\fi
    \expandafter\xdef\csname@s@#1\endcsname{\therunningsection}
    \immediate\write16{\therunningsection. #2}
    \goodbreak\vskip\beforesecskip\noindent%
    \iflabels
          \llap{\tt #1\quad}%
    \fi
    {\sectionfont\the\secnum.\enspace#2}\par\nobreak\noindent\ignorespaces}

\def\subsection#1{\global\advance\subsecnum by 1%
    \xdef\therunningsubsection{\the\subsecnum}%
    \medbreak\smallskip
    \noindent{\sl\the\secnum.\the\subsecnum.\enspace #1}%
    \nobreak\noindent}

\def\section#1#2{\Section{#2}{#1}\par\noindent\ignorespaces}

\def\asection#1{\immediate\write16{#1}%
    \goodbreak\vskip\beforesecskip
    \noindent{\sectionfont#1}\par\nobreak\noindent\ignorespaces}


\def\eqlabel#1{\global\advance\eqnum by 1
    \ifndouble\xdef\anumber{\therunningsection.\the\eqnum}
       \else\xdef\anumber{\the\eqnum}\fi
    \def\usenthrow{1}\ifundefined{@eq@#1}\def\usenthrow{2}\fi
    \expandafter\ifx\csname@eq@#1\endcsname\anumber\def\usenthrow{2}\fi
    \ifodd\usenthrow\immediate\write16
       { --> Possible reference error in \string\eqref{#1} }\fi
    \expandafter\xdef\csname@eq@#1\endcsname{\anumber}
    \ifndouble
       \def\usenthrow{\expandafter\strip\meaning\therunningsection.\the\eqnum}
       \else\def\usenthrow{\the\eqnum}\fi
}

\def\autoeqno#1{\eqlabel{#1}\eqno(\csname@eq@#1\endcsname)
    \iflabels \rlap{\quad\tt #1} \fi
}
\def\autoleqno#1{\eqlabel{#1}\leqno(\csname@eq@#1\endcsname)
    \iflabels \llap{\tt #1 \qquad} \fi
}

\def\therefs{}
\def\bibitem#1#2\par{
    \global\advance\citenum by 1
    \xdef\citation{\the\citenum}
    \def\usenthrow{1}\ifundefined{@c@#1}\def\usenthrow{2}\fi
    \expandafter\ifx\csname@c@#1\endcsname\citation\def\usenthrow{2}\fi
    \ifodd\usenthrow\immediate\write16
      { --> Possible reference error in \string\cref{#1} }\fi
    \expandafter\xdef\csname@c@#1\endcsname{\citation}
\iflabels
     \ifnum\citenum = 1\global\xdef\therefs{\par\noindent\llap{\tt#1\qquad}%
          \ignorespaces#2\par}
     \else 
          \global\xdef\oldrefs{\therefs}
          \global\xdef\therefs{\oldrefs\par\noindent\llap{\tt#1\qquad}%
          \ignorespaces#2\par}
     \fi
\else
     \ifnum\citenum = 1\global\xdef\therefs{\item{[\citation]} #2\par }
     \else 
          \global\xdef\oldrefs{\therefs}
          \global\xdef\therefs{\oldrefs\item{[\citation]} #2\par }%
     \fi
\fi
}


\def\cite#1{\hbox{[\cref{#1}]}}

\newcount\refcount
\refcount=1
\def\listrefs{\frenchspacing
    \asection{References}\par
    \iflabels
          {
          \everypar{\hang\textindent{[\the\refcount]}
          \global\advance\refcount by 1\relax}\therefs
          }
    \else
          \therefs
    \fi
    \nonfrenchspacing}


\def\semi{\hfil\break}

\newdimen\captionwidth
\captionwidth = \hsize
\advance\captionwidth by -2\parindent
\newbox\captionbox

\def\figure#1#2#3{
    \global\advance\fignum by 1
    \xdef\afigure{\the\fignum}
    \def\usenthrow{1}\ifundefined{@f@#1}\def\usenthrow{2}\fi
    \expandafter\ifx\csname@f@#1\endcsname\afigure\def\usenthrow{2}\fi
    \ifodd\usenthrow\immediate\write16
      { --> Possible reference error in \string\figref{#1} }\fi
    \expandafter\xdef\csname@f@#1\endcsname{\afigure}
    \ifnum\fignum = 1\global\xdef\thefigs{\item{Fig.\ \afigure.} #2\ }
    \else%
    \global\xdef\oldfigs{\thefigs}%
    \global\xdef\thefigs{\oldfigs\item{Fig.\ \afigure:} #2\ }%
    \fi%
     \goodbreak\midinsert
     \ifx\epsfbox\undefined
          \immediate\write16{ Fig. \afigure: ignored }
          \noindent\hrule\par\vskip 1cm \noindent\hrule
     \else\immediate\write16{ Fig. \afigure.}
          \center{#3}          
     \fi
     \smallskip
     \setbox\captionbox=\hbox{Figure \afigure: \ignorespaces#2}
     \iflabels
          \ifdim \wd\captionbox < \captionwidth
               \noindent\llap{\tt#1\quad}\centerline{Figure \afigure: \ignorespaces#2}
          \else
               \noindent
               \llap{\tt#1\quad}{\narrower\noindent Figure \afigure: \ignorespaces#2\par}
          \fi
     \else
          \ifdim \wd\captionbox < \captionwidth \centerline{Figure \afigure: \ignorespaces#2}
          \else {\narrower\noindent Figure \afigure: \ignorespaces#2\par}
          \fi
     \fi
     \endinsert
}


\def\table#1#2#3{
    \global\advance\tablenum by 1
    \xdef\atable{\the\tablenum}
    \def\usenthrow{1}\ifundefined{@f@#1}\def\usenthrow{2}\fi
    \expandafter\ifx\csname@f@#1\endcsname\atable\def\usenthrow{2}\fi
    \ifodd\usenthrow\immediate\write16
      { --> Possible reference error in \string\tableref{#1} }\fi
    \expandafter\xdef\csname@f@#1\endcsname{\atable}
    \ifnum\tablenum = 1\global\xdef\thetables{\item{Table \atable.} #2\ }
    \else%
    \global\xdef\oldtables{\thetables}%
    \global\xdef\thetables{\oldtables\item{Table \atable.} #2\ }%
    \fi%
     \goodbreak\midinsert
     \immediate\write16{ Table \atable.}
     \setbox\captionbox=\hbox{Table \atable. \ignorespaces#2}
     \iflabels
          \ifdim \wd\captionbox < \captionwidth
               \noindent\llap{\tt#1\quad}\centerline{Table \atable. \ignorespaces#2}
          \else
               \noindent
               \llap{\tt#1\quad}{\narrower\noindent Table \atable. \ignorespaces#2\par}
          \fi
     \else
          \ifdim \wd\captionbox < \captionwidth \centerline{Table \atable. \ignorespaces#2}
          \else {\narrower\noindent Table \atable. \ignorespaces#2\par}
          \fi
     \fi
     \medskip
     \let\\=\cr
     \centerline{\vbox{\offinterlineskip\halign{\strut\ignorespaces#3}}}
     \medskip
     \endinsert
}


\def\newpage{\vfill\supereject}    
\def\hline{\noalign{\hrule}}
\def\today{\ifcase\month\or January\or February\or
   March\or April\or May\or June\or July\or August\or September\or
   October\or November\or December\fi
   \space\number\day, \number\year}
\def\frac#1#2{{#1\over#2}}

\def\abs#1{\left| #1 \right|}

\def\text#1{{\rm #1}}
\def\degs{\ifmmode {}^\circ \else ${}^\circ$ \fi} 
\def\[{\begingroup$$\let\\=\cr}
\def\]{$$\endgroup\ignorespaces}
\def\\{\hfil\break}
\def\){\hfill\break}

\def\roughly#1{\mathrel{\raise.3ex\hbox{$#1$\kern-.75em\lower1ex%
\hbox{$\sim$}}}}


     \let\gtrsim=\gsim

\def\art#1#2#3#4#5#6{#1, ``#2," {\it#3 \bf#4} (#5) \hbox{#6}.}
\def\artx#1#2#3#4#5#6#7{#1, ``#2," {\it#3 \bf#4} (#5) \hbox{#6}. %
Available from arXiv: {\tt#7}.}
\def\arx#1#2#3{#1, ``#2," arXiv: {\tt#3}.}
\def\proc#1#2#3#4#5{#1, ``#2," #3, {\bf#4}, #5.}
\def\book#1#2#3{#1, {\it#2}, #3.}
\def\ApJ{Astrophys.~J.}

\def\RAN{Izv.\ Ros.\ Akad.\ Nauk, Ser.\ Fiz.}
\def\www#1{\hbox{\tt#1}}
\def\web#1{\hbox{\tt#1}}
\let\thanks=\footnote
\newcount\fnotenum\fnotenum=0
\def\footnote#1{\advance\fnotenum by 1 \thanks{$^{\the\fnotenum}$}{#1}}

\newcount\itemnum
\let\Item=\item
\def\beginenumerate{\itemnum=0\relax \par\begingroup\nobreak
\def\item{\advance\itemnum by 1 \Item{\the\itemnum.}}}
\def\endenumerate{\par\endgroup\medskip}
      
\def\bitem{\item{$\bullet$}}

\def\titleparagraphs{\interlinepenalty=9999
     \leftskip=0.03\hsize plus 0.22\hsize minus 0.03\hsize
     \rightskip=\leftskip \parfillskip=0pt
     \hyphenpenalty=9000 \exhyphenpenalty=9000
     \tolerance=9999 \pretolerance=9000
     \spaceskip=0.333em \xspaceskip=0.5em }
\def\center#1{\par{
     \def\\{\break} \titleparagraphs \noindent #1\par}}

\def\preprint#1{\par\rightline{\tt #1}\bigskip}

\def\remark{\global\advance\remarknum by 1
   \smallskip{\it Remark \the\remarknum.}\enspace\ignorespaces}

\font\titlefont = cmbx10 at 12 pt
\def\title#1{\center{\baselineskip=14pt\titlefont\ignorespaces#1}}
\def\author#1{\bigskip\center{\ignorespaces\bf#1}}
\def\address#1{\smallskip\center{\ignorespaces#1}}
\def\date#1{\medskip\centerline{#1}}


\input epsf
\let\label=\autoeqno
\let\ref=\eqref

\def\fig#1{\epsfxsize=0.49\hsize \epsfbox{#1.eps}}
\def\bfig#1{\epsfysize=0.355\vsize \epsfbox{#1.eps}}

\hyphenation{an-a-lyse be-hav-iour}
\def\Ne{N_{\rm e}}
\def\Nmin{N_{\rm min}}
\bibitem{Hess}
   V.F. Hess, {\it Phys. Z.} {\bf13} (1912) 1084.
   
\bibitem{Chili03}
   \art{A. Chilingarian et al.}{Detection of the high-energy cosmic rays 
   from the Monogem Ring}{Astrophys.~J.}{597}{2003}{L129}

\bibitem{KASCADE}
   \artx{T. Antoni et al. for the KASCADE Collaboration}{Large scale 
   cosmic-ray anisotropy with KASCADE}{\ApJ}{604}{2004}{687}
   {astro-ph/0312375}\\
   \artx{K.-H. Kampert et al.}{Cosmic rays in the 
   `knee'-region. Recent results from KASCADE}{Acta Phys. Polon.}{B35}
   {2004}{1799}{astro-ph/0405608}



  
\bibitem{Nesterova}
   \artx{G. Benk\'o et al.}{Search for Sources of Primary Cosmic Rays 
   at Energies Above 0.1 Pev at Tien Shan}{Izv. RAN, Ser. Fiz.}{69}
   {2004}{1599}{astro-ph/0502065}

\bibitem{Andyrchi}
   \arx{V.A. Kozyarivsky et al.}{Mean diurnal variations of cosmic ray
   intensity as measured by Andyrchi Air Shower Array}{astro-ph/0406059}


\bibitem{Clay}
   \proc{R.W. Clay et al.}{Anisotropies between $10^{14}$~eV and
   $10^{18}$~eV}{Proc. 25th ICRC, Durban, 1997}{4}{185}

\bibitem{Kulikov}
   G.V. Kulikov, G.B. Khristiansen, \it Zh. Eksp. Teor. 
   Fiz. \bf35 \rm(1958) 635.

\bibitem{Roulet}
   \artx{E. Roulet}{Astroparticle Theory: Some New Insights into High 
   Energy Cosmic Rays}{Int.~J. Mod. Phys.}{A19}{2004}{1133}{astro-ph/0310367}

\bibitem{Wick}
   \artx{S.D. Wick, C.D. Dermer, A. Atoyan}{High-Energy Cosmic Rays from
   Gamma-Ray Bursts}{Astropart. Phys.}{21}{2004}{125}{astro-ph/0310667}

\bibitem{Dar}
   \arx{A. Dar}{The Origin of Cosmic Rays---A~96-Year-Old Puzzle Solved?}
   {astro-ph/0408310}

\bibitem{Muraishi}
   \arx{H. Muraishi, S. Yanagita, T. Yoshida}{Galactic modulation of 
   extragalactic cosmic rays: Possible origin of the knee in the cosmic
   ray spectrum}{astro-ph/0502132; v.2}

\bibitem{Gaisser01}
   \artx{T.K. Gaisser}{Origin of cosmic radiation}{AIP Conf. 
   Proc.}{558}{2001}{27}{astro-ph/0011524}

\bibitem{EW-JPG}
   \art{A.D. Erlykin, A.W. Wolfendale}
   {A single source of cosmic rays in the range 10$^{15}$--10$^{16}$~eV}
   {J. Phys.~G: Nucl.\ Part.\ Phys.}{23}{1997}{979}\semi
   \art{---}{Structure in the cosmic ray spectrum: an update}
   {J. Phys.~G: Nucl.\ Part.\ Phys.}{27}{2001}{1005}

\bibitem{Thorsett}
   \artx{S.E. Thorsett et al.}{Pulsar PSR B0656+14, the Monogem Ring,
   and the origin of the `knee' in the primary cosmic ray
   spectrum}{Astrophys.~J.}{592}{2003}{L71}{astro-ph/0306462}


\bibitem{AGASA}
   \artx{N. Hayashida et al.}{Updated AGASA event list above 
   $4\times10^{19}$~eV}{Astrophys.~J.}{522}{1999}{225}{astro-ph/0008102}

\bibitem{Fomin99}
   \proc{Yu.A. Fomin et al.}{New results of the EAS-1000 Prototype
   operation}{Proc. 26th ICRC, Salt Lake City, 1999}{1}{286}

\bibitem{Zhengzhou}
   \proc{L. Sun, S. Sun}{The cosmic ray incident directions observed
   from the Northern and Southern hemispheres}{Proc. 25th ICRC,
   Durban, 1997}{4}{165}

\bibitem{Octave}
   \book{J.W. Eaton}{GNU Octave: A High-Level Interactive   
   Language for Numerical Computations}{Edition~3 for version 2.0.13, 1997}
   (\www{http://www.octave.org/})

\bibitem{Green}
   D.A. Green, ``A Catalogue of Galactic Supernova Remnants
   (2004 January version),'' Mullard Radio Astronomy Observatory,
   Cavendish Laboratory, Cambridge, United Kingdom\\ 
   (available at \www{http://www.mrao.cam.ac.uk/surveys/snrs/});
   see also \artx{D. A. Green}{Galactic Supernova Remnants: 
   an~Updated Catalogue and Some Statistics}{Bull. Astron. Soc.
   India}{32}{2004}{335}{astro-ph/0411083}


\bibitem{Volk03}
   \arx{H.J V\"olk}{TeV gamma-ray observations and the origin
   of cosmic rays~III}{astro-ph/0312585}
   
\bibitem{BV2004}
   \arx{E.G. Berezhko, H.J V\"olk}{Direct evidence of efficient cosmic 
   ray acceleration and magnetic field amplification in Cassiopeia~A}
   {astro-ph/0404203}
   
\bibitem{CR-Japan}
   \proc{T. Kobayashi et al.}{High energy cosmic-ray electrons beyond
   100~GeV}{Proc. 26th ICRC, Salt Lake City, 1999}{3}{61}\semi
   \proc{K. Yoshida et al.}{The origin of high energy cosmic-ray
   electrons and nearby supernova remnants}{Proc. 28th ICRC, Tsukuba,
   2003}{1}{1993}

\bibitem{EW01}
   \arx{A.D. Erlykin, A.W. Wolfendale}{The origin of the knee in the 
   cosmic-ray energy spectrum}{astro-ph/0103477}

\bibitem{ATNF}
   The ATNF Pulsar Database,
   \www{http://www.atnf.csiro.au/research/pulsar/psrcat/}; see also:\\
   \arx{R.N.~Manchester, G.B.~Hobbs, A.~Teoh, M.~Hobbs}{The ATNF Pulsar
   Catalogue}{astro-ph/0412641}

\bibitem{TaylorCordes}
   J.H. Taylor, J.M. Cordes, \it Astrophys.~J. \bf411 \rm(1993) 674.

\bibitem{EW04}
   \artx{A.D. Erlykin, A.W. Wolfendale}{Cosmic rays and the Monogem 
   supernova remnant}{Astropart. Phys.}{22}{2004}{47}{astro-ph/0404530}

\bibitem{Simbad}
   The SIMBAD database, \web{http://simbad.u-strasbg.fr/Simbad}

\bibitem{Parizot}
   \art{E. Parizot}{Superbubbles \& the Galactic evolution of $^6$Li,
   Be and~B}{Astron. Astrophys.}{362}{2000}{786}\\  
   \arx{E. Parizot, A. Marcowith, E. van der Swaluw, A.M.~Bykov,
   V.~Tatischeff}{Superbubbles and Energetic Particles in the Galaxy.
   I:~Collective effects of particle acceleration}{astro-ph/0405531}\\
   \arx{E. Parizot}{Cosmic-rays: an unsolved mystery at all
   energies!}{astro-ph/0501274}

\bibitem{galH2}
   \art{R. Paladini et al.}{A radio catalog of Galactic HII regions for
   applications from decimeter to millimeter wavelengths}{Astron. \&
   Astrophys.}{397}{2003}{213}

\bibitem{IzvRAN01}
   \art{O.V. Vedeneev et al.}{EAS clusters with 
   $\Ne\sim5\times10^4$}{\RAN}{65}{2001}{1674}

\bibitem{preprint}
   Yu.A. Fomin et al., ``Clusters of EAS with electron number 
   $\gtrsim10^4$,'' Preprint SINP MSU 2002-9/693. Available from arXiv:
   {\tt astro-ph/0203478}.

\bibitem{HiRes}
   \arx{R.U. Abbasi, T. Abu-Zayyad, J.F. Amann et al.}{Search for Point
   Sources of Ultra-High Energy Cosmic Rays Above 40 EeV Using a Maximum
   Likelihood Ratio Test}{astro-ph/0412617}


\bibitem{Tarle}
   \proc{G. Tarl\'e et al.}{Limitations of the transport of cosmic ray 
   antimatter from distant galaxies}{Proc. 25th ICRC, Durban, 
   1997}{4}{205}

\bibitem{VV2003c}
   \art{M.P. Veron-Cetty, P. Veron}{Quasars and Active Galactic 
   Nuclei}{Astron.~\& Astrophys.}{412}{2003}{399}

\bibitem{APG}
   \art{H. Arp}{Atlas of peculiar galaxies}{\ApJ, Suppl. 
   Ser.}{14}{1966}{1}
   
\bibitem{NED}
   The NASA/IPAC Extragalactic Database,
   \web{http://nedwww.ipac.caltech.edu/index.html}
   
\bibitem{AGASA99}
   \artx{M. Takeda et al.}{Small-scale anisotropy of cosmic rays above
   10$^{19}$~eV observed with the Akeno Giant Air Shower
   Array}{\ApJ}{522}{1999}{225}{astro-ph/9902239}

\bibitem{Brunthaler}
   \artx{A. Brunthaler, M.J. Reid, H. Falcke, L.J. Greenhill, C. Henkel}{The
   Geometric Distance and Proper Motion of the Triangulum Galaxy
   (M33)}{Science}{307}{2005}{1440}{astro-ph/0503058}

\bibitem{M33}
   \arx{G. Dubus et al.}{High resolution Chandra X-ray
   imaging of the nucleus of M33}{astro-ph/0406310}

\bibitem{LAAS}
   \art{N. Ochi, T. Wada, Y. Yamashita et al.}{Anisotropy of
   Successive Air Showers}{Nucl. Phys. B (Proc. Suppl.)}{97}{2001}{173}

\bibitem{Milagro}
   \arx{P.M. Saz Parkinson for the Milagro Collaboration}{Detection of
   TeV gamma-rays from extended sources with Milagro}{astro-ph/0503244}

\bibitem{EWW01}
   \art{A.D. Erlykin, T. Wibig, A.W. Wolfendale}{A universal origin for
   cosmic rays above 10$^7$~GeV?}{New Journal of Physics}{3}{2001}{18.1}
   (\www{http://www.njp.org/})

\bibitem{SEDS}
  Students for the Exploration and Development of Space (SEDS),
  available at \www{http://www.seds.org/}.

\preprint{astro-ph/0407138; v.2}

\title{
   A Search for Outstanding Sources of PeV Cosmic Rays:\\
   Cassiopeia~A, the Crab Nebula, the Monogem Ring~--- \\
   But How About M33 and the Virgo Cluster?}

\author{G. V. Kulikov, M. Yu.\ Zotov}
\address{
   D. V. Skobeltsyn Institute of Nuclear Physics\\
   Moscow State University, Moscow 119992, Russia\\
   \tt $\{$kulikov,zotov$\}$@eas.sinp.msu.ru}

\date{April 22, 2005}

\bigskip
\centerline{\bf Abstract}
{\narrower\noindent
   We study arrival directions of $1.4\cdot10^6$ extensive air showers
   (EAS) registered with the EAS--1000 Prototype Array in the energy 
   range 0.1--10~PeV.
   By applying an iterative algorithm that provides uniform distribution
   of the data with respect to sidereal time and azimuthal angles,
   we find a number of zones with excessive flux of cosmic rays (CRs)
   at $\ge3\sigma$ level.
   We compare locations of the zones with positions of galactic
   supernova remnants (SNRs), pulsars, open star clusters,
   OB-associations, and regions  of ionized hydrogen and find remarkable
   coincidences, which may witness in favour of the hypothesis that
   certain objects of these types, including  the SNRs Cassiopeia~A, the
   Crab Nebula, the Monogem Ring and some other, provide a noticeable
   contribution to the flux of  CRs in the PeV range of energies.
   In addition, we find certain signs of a contribution from the M33 galaxy
   and a number of comparatively nearby groups of active galactic nuclei
   and interacting galaxies, in particular those in the Virgo cluster of
   galaxies.
   The results also provide some hints for a search of possible sources
   of ultra-high energy (UHE) cosmic rays and support an earlier idea
   that a part of both UHE and PeV CRs may originate from the same
   astrophysical objects.

}

\bigskip
\bigskip
\hfill{\it TMTOWTDI}\thanks{$^*$}{There's More Than One Way To Do It.}

\hfill{The Perl community mantra}

%

\section{Introduction}{sec:intro}
   In spite of the fact that cosmic rays were discovered more than
   90 years ago~\cite{Hess}, the problem of their origin for energies 
   greater than 100~TeV remains unsolved.
   One of the important directions in numerous approaches to the problem
   is an analysis of arrival directions of CRs.
   Such an analysis has been performed with the data sets obtained with
   practically all extensive air shower experiments,
   see~\cite{Chili03}, \cite{KASCADE}, \cite{Nesterova},
   and~\cite{Andyrchi} for the latest reports and a list of references.
   One of the main results of the majority of these investigations is 
   that there is no significant anisotropy in the energy range 
   0.1--100~PeV~\cite{Clay}.
   
   Another intriguing and long-standing problem in cosmic ray physics 
   is the so called `knee' around $3\cdot10^{15}$~eV in the 
   CR energy spectrum~\cite{Kulikov}.
   The knee is a point where the spectral index of the all-particle
   differential power-law spectrum changes from approximately~$-2.7$ 
   to~$-3.1$.
   There are a number of models aimed to explain this feature (as well
   as the origin of CRs), see, e.g.,~\cite{Roulet}, \cite{Wick}, \cite{Dar},
   and~\cite{Muraishi} for a number of recent works and~\cite{Gaisser01}
   for an earlier review.
   Still, none of them seems to be fully established yet.
   One of the modern `astrophysical' approaches to the problem is
   the `single source model' by Erlykin and Wolfendale~\cite{EW-JPG}.
   The model explains the knee as a result of a contribution from one
   comparatively recent ($\sim100$~kyr) and nearby ($\sim300$~pc)
   supernova remnant accompanied by a pulsar.
   Evidently, an analysis of arrival directions of CRs in the energy
   range around the knee can be useful as a (partial) experimental test
   of the model.

   An interest to the single source model has recently received an
   additional impulse after the discovery that the pulsar PSR
   B0656+14 is located near the center of the Monogem Ring SNR
   at the distance of about 290~pc~\cite{Thorsett}.
   Since then, there has been a number of reports on the
   analysis of the flux of cosmic rays with energies around the knee
   from the direction to this source.
   Interestingly, some of the results are controversial.
   Namely, Chilingarian et al.~\cite{Chili03} and Benk\'o et
   al.~\cite{Nesterova} find a significant excess of EAS from zones near
   the pulsar PSR B0656+14 while the KASCADE collaboration considers
   this excess as negligible~\cite{KASCADE}.
   To shed an additional light on the situation,  we perform an analysis
   of arrival directions of EAS registered with the EAS--1000 Prototype
   Array, which has been operating at Moscow State University.
   Unfortunately, the energy range covered by the array is not
   sufficient for a comprehensive investigation of arrival directions
   of EAS around the knee since it mostly covers a range of energies
   just below the knee.
   Still, we find the first results of this work sufficiently interesting 
   and promising to report.

   The most important result is the discovery of a whole number of
   zones with an excessive flux (ZEF) of CRs around 1~PeV at $\ge3\sigma$
   level.
   The majority of the ZEF correspond to the locations of galactic SNRs,
   pulsars,
   open star clusters, OB-associations, and regions of ionized hydrogen.
   In particular,
   we confirm the conclusions of Chilingarian et al.~\cite{Chili03}
   and Benk\'o et al.~\cite{Nesterova} on the
   excessive flux of CRs from the direction to the Monogem Ring.
   Still, a number of the zones have no or just a few objects of the above
   types nearby.
   We find that one of these `empty' or `underfilled' ZEF contains
   the M33 galaxy inside, while the majority of others have 
   neighbouring active galactic nuclei (AGN) and/or interacting galaxies
   at redshifts $z<0.01$ with a big group of them located in the
   Virgo cluster of galaxies.
   In addition, we
   find a few remarkable coincidences between positions of the ZEF
   and arrival directions of ultra-high energy cosmic rays registered
   with the AGASA array~\cite{AGASA}.

\section{Experimental Data}{sec:data}   
   The EAS--1000 Prototype Array consists of eight detectors 
   situated in the central part of the EAS MSU array along longer sides 
   of the $64\,{\rm m}\times22\,{\rm m}$ rectangle~\cite{Fomin99}.
   The array is located at $37^\circ32.5'$E, $55^\circ41.9'$N 
   at approximately 200~m above sea level.

   The data set under consideration includes 1,668,489 EAS registered
   during 203 days of operation of the array in the period from
   August~30, 1997, till February~1, 1999.
   The arrival directions are determined for 1,366,010 EAS.
   A number~$\Ne$ of charged particles (electrons) is found for 826,921
   EAS.
   It happens that~$\Ne$ of 95.2\% of showers with zenith angles
   $\theta\le45^\circ$ lie in the range $10^4$--$1.1\cdot10^6$ particles
   with $\bar \Ne=1.2\cdot10^5$ and the median value equal to
   $6.0\cdot10^4$.
   Thus we estimate that the overwhelming majority of primary cosmic
   rays that give life to the EAS in the data set have energies in the
   range $E\approx0.1$--10~PeV.
   Still, only 92,212 EAS have $\Ne>3\cdot10^5$.
   Thus one may treat the data set as mostly covering an interval
   of energies just below the knee.

\section{Method of Data Analysis}{sec:method}
   Perhaps the majority of investigations of anisotropy  rely on the
   Rayleigh method for the calculation of the amplitude and the
   phase of the first and (sometimes) the second harmonics~\cite{Clay}.
   The method by itself is based on the Fourier transform. 
   This does not seem to be a fruitful approach in all situations.
   For example, one can easily check that an analysis of the
   one-dimensional Fourier spectrum does not necessarily give reliable
   results if the amplitude of a periodic signal is less than or of the
   order of 0.3\% of that of noise for a sample that  consists of $10^6$
   points.
   Thus we expect that a similar situation can take place in the case of
   anisotropy analysis.
   This makes us develop another approach, which may be considered
   as an advanced version of the method employed in~\cite{Zhengzhou}.
   The algorithm consists of three main steps: data alignment,
   data averaging, and selection of ZEF.

\smallskip
   {\bf Step 1: Data Alignment.}
   The main aim of the step is to obtain a data set that consists
   of EAS with a uniform distribution of azimuthal angles~$\phi$ and
   sidereal time~$s$.

\beginenumerate   
\item
   Fix values $\Delta\phi$ and $\Delta s$ of the width 
   of bins in histograms of~$\phi$ and~$s$.
   Produce a histogram of~$\phi$.
   By~$N_i$ denote the number of EAS in the $i$th bin.
   
\item
   In the histogram of~$\phi$, find a bin with the least number of
   showers,~$\Nmin$.
   At random, exclude $(N_i - \Nmin)$ showers from each bin.
   
\item
   Obtain a histogram of~$s$ for the remaining data set.
   With this histogram, perform an `alignment' of data exactly as
   in item~2.
   
\item
   Produce a histogram of~$\phi$ for the data set obtained in item~3.
   Finish the procedure if 
\[
   \max_i \abs{N_i - \bar N} \le 2.5, \quad \hbox{\rm and} \quad
   \max_i \abs{N_i - \bar N} < 3\sigma,
   \label{stop.align}
\]
   where $\bar N$ is the mean number of EAS in the bins of the histogram,
   and~$\sigma$ is the standard deviation.
   Otherwise, repeat items~2 and~3.
\endenumerate
   
   Conditions (\ref{stop.align}) were chosen after comparing a number
   of empirical criteria.
   In practice, the second inequality is always fulfilled earlier than
   the first one and may be omitted.
   
   We stress that the data set obtained as a result of Step~1 contains
   equal number of EAS in each bin of the histogram of the sidereal
   time~$s$ thus providing uniform distribution of the data
   with respect to~$s$ at the chosen scale~$\Delta s$.
   The same is approximately true for the distribution with respect
   to~$\phi$.

   For the data set under consideration, Step~1 leads to an exclusion
   of up to~20\% of all EAS in the data set.
   It is obvious that results of an analysis of the `aligned' data set
   will strongly depend on the choice of the excluded EAS.
   Hence we come to the necessity to implement Step~1 numerous times
   thus obtaining a family of `aligned' data sets.
   Each time, an independent sample of EAS is excluded from the whole
   data set.
   
\smallskip
   {\bf Step 2: Data Averaging.}
   For each `aligned' data set, produce a map of the distribution of
   EAS arrival directions in equatorial coordinates.
   Each map consists of $1^\circ\times1^\circ$ cells (`unit' cells)
   and covers the complete observable region with the right ascension 
   $\alpha=0\dots24$~h and the declination $\delta>-13^\circ$.
   For each cell in all maps, a number of EAS inside the cell is found.
   After this, an averaged map is produced. 
   
   At this point, we are ready to join unit cells into larger ones and
   look for cells with an excessive flux (CEF) of cosmic rays.
   We do not analyse unit cells because it is seems unlikely that charged 
   particles of PeV energies preserve directions to their sources with such
   precision.
   
\smallskip
   {\bf Step 3: Selection of CEF.}

\beginenumerate
\item
   Divide the averaged map into strips of equal width 
   $\Delta\delta\ge3^\circ$ in the declination.
   The boundaries of strips of equal width are shifted with
   respect to each other by~$1^\circ$.
   
\item
   Split each strip into cells of equal width~$\Delta\alpha$.
   For each strip and each partition into cells, find the mean
   number~$\bar N$ of EAS for all cells within the strip
   and the standard deviation~$\sigma$.
   Mark a cell with~$N^*$ EAS inside as having an excessive CR flux if
   $N^*-\bar{N}\ge3.1\sigma$, and $N^*\ge10$.
\endenumerate

\remark
   It often happens that the algorithm selects a number of intersecting
   or overlapping CEF, so that it is inconvenient to discuss each of
   them separately.
   Instead, we shall discuss {\it zones\/} of excessive flux (ZEF), by 
   which we mean sets of intersecting or overlapping CEF.
   For simplicity, we shall use this term even in case a ZEF consists 
   of a single cell.

\remark
   It is not important for the algorithm how the values of~$\Delta\alpha$  
   and~$\Delta\delta$ are chosen.
   The majority of ZEF discussed below are combined of cells such
   that $\Delta\alpha$ equals a rounded value of
   $\Delta\delta/\cos\bar{\delta}$, where~$\bar{\delta}$ is the mean
   value of the declination for the strip.
   This guarantees that cells with the same~$\Delta\delta$ but located
   at different declination, have an approximately equal area.
   This is a traditional choice for similar investigations.
   In what follows,we call these cells `regular'.
   Still, there are no reasons to restrict the analysis to regular cells
   only.
   Thus, we have also studied cells with arbitrary values of~$\Delta\alpha$
   providing that they are integer dividers of $360^\circ$.
   These `irregular' cells are in no sense worse than `regular' ones.
   More than this, in some cases an employment of irregular cells
   allows one to find CEF that are more pronounced than their regular
   counterparts in the sense of deviation from~$\bar{N}$.
   On the other hand, for any regular CEF, there is an irregular
   one with the same or close size.
   A number of interesting irregular CEF will be presented below.

   It is important to mention that all CEF have a neighbourhood with
   a flux of CRs that is higher than the average for the strip though
   it does not reach the level of~$3.1\sigma$.
   In our opinion, this means that CEF are not just random fluctuations
   of the level of the flux caused by the algorithm. 
   
\remark
   The rule used to choose CEF implicitly assumes that the number of EAS
   within cells of any strip obey the Gaussian distribution with the
   corresponding parameters.
   The hypothesis seems to be possible since these numbers are 
   generically non-integer due to the averaging of the data set.
   Nevertheless, one may choose another rule to select CEF.

\remark
   An analysis of the algorithm described above has revealed that maps
   of `averaged' EAS arrival directions obtained with 100 cycles of
   alignment differ very little from each other.
   Still, in order to make our conclusions even more robust, we present
   results obtained for 1000 cycles.
   To perform calculations, we have employed GNU Octave~\cite{Octave}
   running in Linux.

\remark
   In what follows, we present results obtained for
   $\Delta\phi=20^\circ$ and $\Delta s=3$~h, which provide a
   comparatively small number of excluded EAS and an acceptable speed of
   the algorithm.
   An analysis of averaged maps obtained for different values
   of~$\Delta\phi$ and~$\Delta s$ in the range
   $\Delta\phi=10^\circ\dots30^\circ$ and $\Delta s=1\dots3$~h
   and cells of the size $5^\circ\times5^\circ$ has revealed that
   the results differ from those obtained for
   $\Delta\phi=20^\circ$ and $\Delta s=3$~h by at most~$0.1\sigma$.
   Thus one may expect that the condition $N^*-\bar{N}\ge3.1\sigma$ used
   to select CEF guarantees that $N^*-\bar{N}\ge3.0\sigma$ for other
   sensible values of~$\Delta\phi$, $\Delta s$, and sizes of cells.

\figure{fig:SNRs}{
   Zones with excessive flux of cosmic rays (rectangles)
   and galactic SNRs~($\circ$) in the region $\delta\ge-15^\circ$.
   The ZEF are numbered from top to bottom according to the
   values of their upper boundaries. 
   The curves show the galactic plane.
   Filled circles show SNRs located within $2^\circ$-neigh\-bourhoods
   of the ZEF.}{\bfig{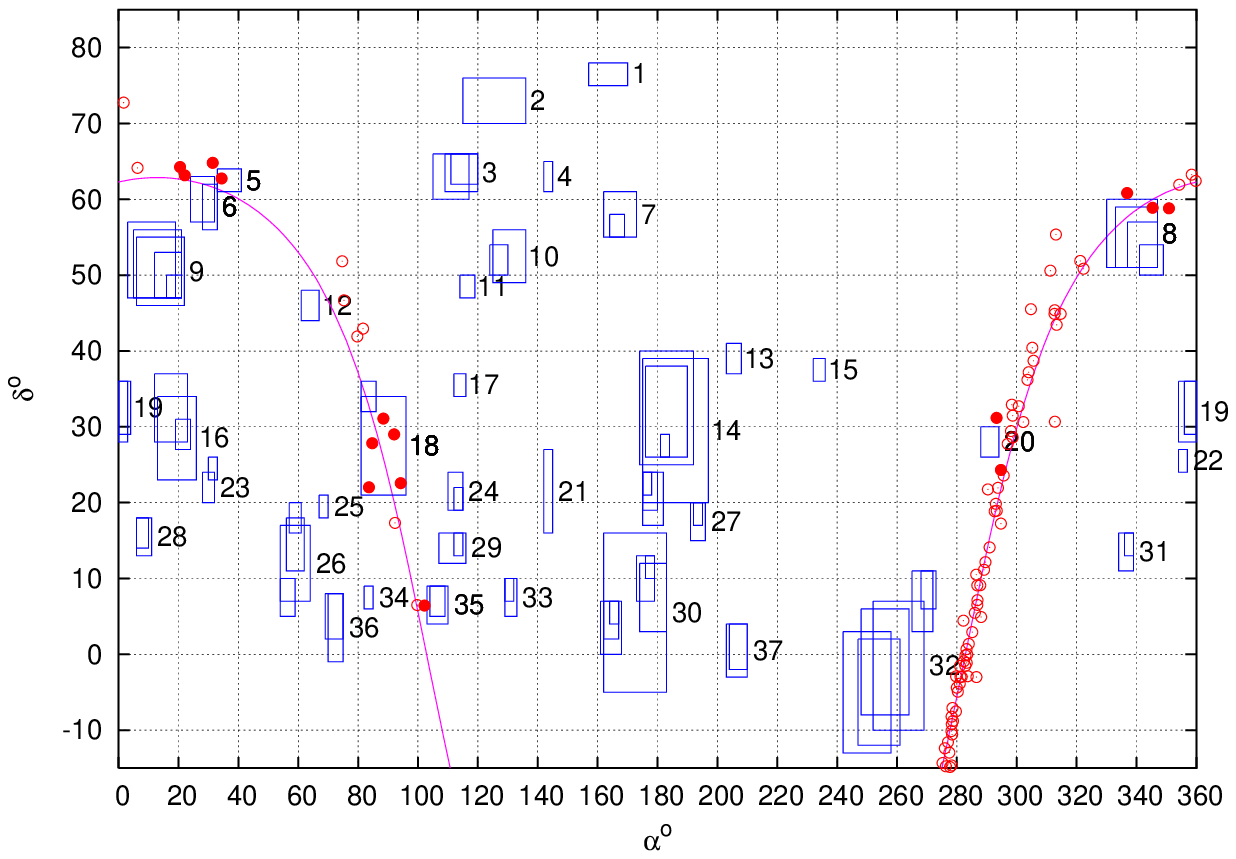}}

\section{Analysis of the Data Set}{sec:whole}
   The above procedure applied to the whole data set resulted in
   the `averaged' set of 1,134,325 EAS and 193 regular and an even 
   greater number of irregular CEF.
   The regular CEF and a number of irregular CEF were joined into 37
   ZEF, see \Fig{fig:SNRs}.
   A number of CEF are embedded into each other or shifted with respect to
   each other by~$1^\circ$.
   To make this and the following figures as clear as possible, we omit
   some of these CEF but show their joint boundaries instead.
   
   An important question is how many EAS are there inside ZEF?
   Below we shall give exact numbers but it may be useful to remember
   the following: (i)~any CEF contains at least 10 EAS;
   (ii)~any unit cell located in the region
   $26^\circ\le\delta\le80^\circ$ contains $\gtrsim10$ EAS;
   (iii)~the mean number of EAS inside unit cells located in the region 
   $19^\circ\le\delta\le85^\circ$ is greater than~9;
   (iv)~$\min\delta=-12.38^\circ$ but there are only 50 EAS with
   $\delta<0^\circ$.

\subsection{Galactic SNRs.}
   Let us begin an analysis of positions of possible CR sources with
   galactic supernova remnants.
   For this purpose, we use the January 2004 version of the
   D.~Green Catalogue~\cite{Green}.

   \Figure{fig:SNRs} shows positions of 96~galactic SNRs located in the
   region $\delta\ge-15^\circ$.
   (The Monogem Ring remnant is not shown since it is not present in the
   main list of~\cite{Green}.)
   It happens that 15~SNRs lie inside or at angular distances
   $\Delta\le2^\circ$ from the nearest ZEF.
   Namely, 
\bitem
   ZEF No.~5 has the SNR G132.7+1.3 (HB3) inside and the remnant
   G130.7+3.1 (SN1181) located at the angular distance $\Delta=1.1^\circ$;
\bitem
   ZEF No.~6 has the SNR G127.1+0.5 (R5) at $\Delta=0.9^\circ$,
   the SNR G126.2+1.6 at $\Delta=2.0^\circ$, and the remnants associated with
   ZEF No.~5 at $\Delta=1.0^\circ$ and $\Delta=1.8^\circ$ respectively;
\bitem   
   ZEF No.~8 has the SNRs G109.1$-$1.0 (CTB~109) inside,
   G106.3+2.7 at $\Delta=0.8^\circ$, and G111.7$-$2.1 (Cassiopeia~A) at
   the angular distance $\Delta=2.0^\circ$;
\bitem
   ZEF No.~18 has five SNRs inside: G179.0+2.6, G180.0$-$1.7 (S147),
   G182.4+4.3, G184.6$-$5.8 (the Crab Nebula, SN1054), and G189.1+3.0 (IC443);
\bitem
   ZEF No.~20 has the SNR G65.3+5.7, located at $\Delta=1.2^\circ$ and
   the SNR G59.8+1.2 at $\Delta=1.8^\circ$.
\bitem   
   ZEF No.~35 has the SNR G206.9+2.3 (PKS~0646+06), located at 
   $\Delta=0.8^\circ$.
   
\table{tab:SNRcells}{
   Some parameters of the ZEF with galactic SNRs inside or at
   angular distances $\Delta\le2^\circ$.
   Notation by columns:
   `ZEF' is the number of the ZEF as shown in \Fig{fig:SNRs}
   (letters `Ir' mark irregular CEF),
   $\alpha^\circ$ and~$\delta^\circ$ are the ranges of values of the 
   right ascension and the declination for the listed cells respectively,
   $N^*$ is the number of EAS in the cell,
   $\bar{N}$ is the mean number of EAS in the corresponding strip
   for the given value of~$\Delta\alpha$,
   $\sigma$ is the standard deviation for the strip,
   $D=(N^*-\bar{N})/\sigma$, i.e., the deviation of~$N^*$ from~$\bar{N}$
   in units of~$\sigma$.}{
\vrule \hfil$\,\,$#          &        
\vrule \hfil$\;\,#\,$        &        
\vrule \hfil$\;\,#\,\,$\hfil &        
\vrule \hfil$\;\,#\,\,$      &        
\vrule \hfil$\;\;#\,\,$      &        
\vrule \hfil$\;\,#\,\,$      &        
\vrule \hfil$\;#\;$\hfil              
\vrule \\ 
\hline
\vphantom{$\sqrt A$}
 ZEF&\alpha^\circ\quad\;&\delta^\circ&N^*\;\,&\bar{N}\quad&\sigma\quad&D\\
\hline
 5 (Ir) &  33\dots 41 &  61\dots64 & 1673.2 & 1546.5 & 39.10 &  3.24 \\
\hline
 6 (Ir) &  24\dots 32 &  57\dots63 & 3467.0 & 3297.0 & 54.09 &  3.14 \\
        &  29\dots 32 &  59\dots62 &  703.6 &  615.8 & 23.31 &  3.76 \\
        &  29\dots 33 &  56\dots62 & 1804.2 & 1680.8 & 38.03 &  3.24 \\
\hline
 8      & 331\dots347 &  51\dots60 & 10883.4& 10546.1& 105.47&  3.20 \\
        & 342\dots348 &  50\dots54 & 1969.3 & 1818.6 & 41.45 &  3.64 \\
\hline
18      &  81\dots 86 &  32\dots36 &  984.9 &  901.8 & 26.69 &  3.11 \\
        &  81\dots 96 &  21\dots34 & 5445.8 & 5266.3 & 57.48 &  3.12 \\
\hline
20 (Ir) & 288\dots294 &  26\dots30 &  734.5 &  660.7 & 21.72 &  3.39 \\
\hline
35      & 103\dots108 &   4\dots 9 &  24.6  &  14.0  &  3.38 &  3.14 \\
        & 104\dots108 &   5\dots 9 &  20.5  &  10.0  &  3.14 &  3.34 \\
\hline
}

   Notice that a number of these SNRs are widely discussed
   in literature as possible sources of CRs in the region of the knee.
   Among them, Cassiopeia~A (see, e.g., \cite{Volk03}, \cite{BV2004} and
   references therein), the Crab Nebula, and the SNRs S147 and
   G65.3+5.7~\cite{CR-Japan}, \cite{EW01}.
   
   An extension of the selection zone up to the angular distance
   $\Delta\le3^\circ$ adds two more SNRs to the list, namely
   G59.5+0.1 and G63.7+1.1 both lying near ZEF No.~20.
   Another prominent SNR, the Monoceros Nebula, is located
   at $\Delta=3.2^\circ$ to the west from ZEF No.~35.

   \Tab{tab:SNRcells} contains some parameters of the discussed ZEF.
   Among them, ZEF No.~5, 6, and~20 are made of irregular cells.
   In the boundaries shown in \Fig{fig:SNRs}, ZEF No.~6, 8 and~35 consist 
   of 18, 10, and 4 CEF respectively.
   For the sake of brevity, only the most interesting of them are
   presented in \Tab{tab:SNRcells}.
   Notice that all these ZEF except No.~35 lie in the region with
   a high density of EAS per unit cell and thus contain hundreds and
   even thousands of showers.

\figure{fig:PSRs}{
   The ZEF and pulsars in the region $\delta\ge-15^\circ$~\cite{ATNF}.
   Different symbols denote pulsars located at different distances~$d$
   from the solar system:
   $\circ\,$--- $d\le3$~kpc, $\triangle\,$--- $3<d\le6$~kpc, 
   $\times$--- $d>6$~kpc or the distance is unknown.
   Filled circles, filled triangles, and asterisks~($*$) are used
   respectively to show pulsars located within $2^\circ$-neigh\-bourhoods
   of the ZEF.}{\bfig{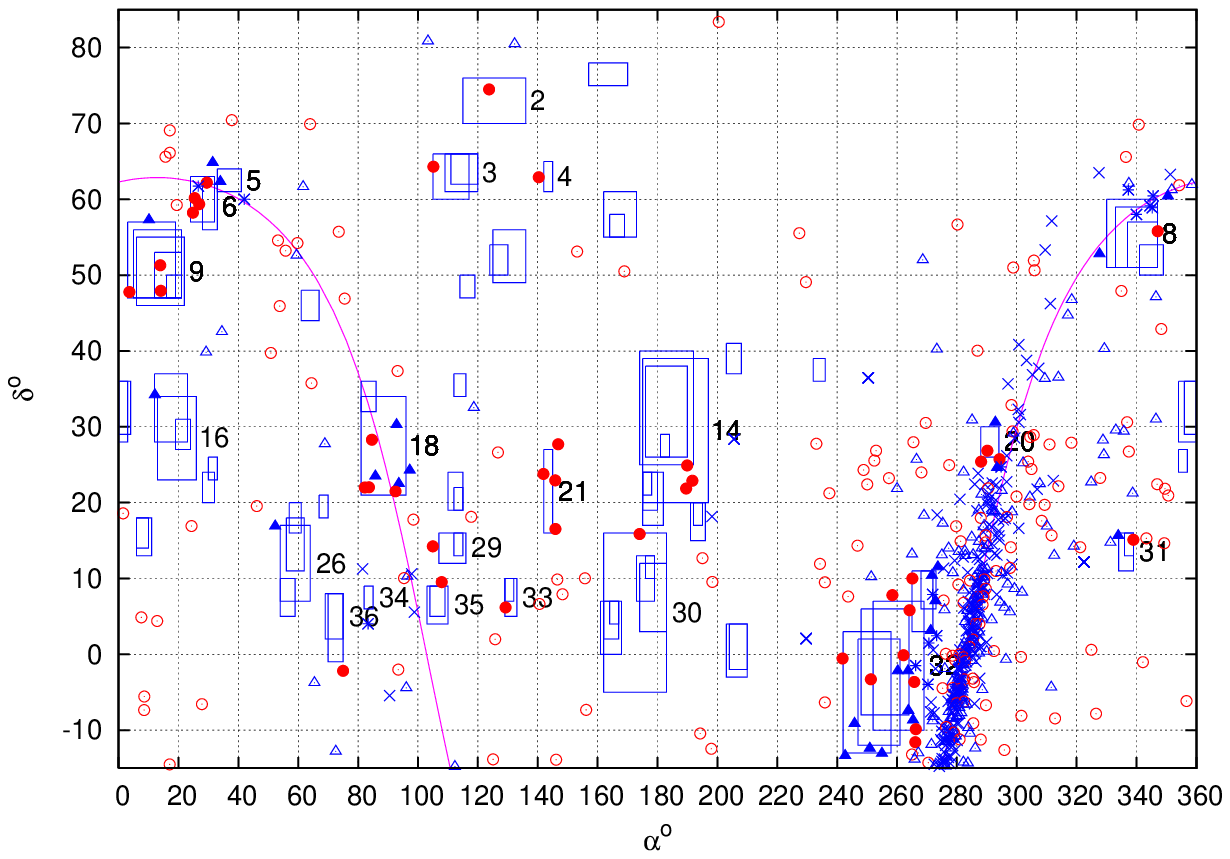}}

\table{tab:PSRinside}{
   Some parameters of the ZEF with pulsars inside.
   For composite ZEF made of multiple CEF, the corresponding number is
   given in parenthesis after the number of the zone.
   See \Tab{tab:SNRcells} for other notation and parameters of ZEF
   No.~5, 6, 8, 18, and~20.}{
\vrule \hfil$\,\,$#          &        
\vrule \hfil$\;\,#\,$        &        
\vrule \hfil$\;\,#\,\,$\hfil &        
\vrule \hfil$\;\,#\,\,$      &        
\vrule \hfil$\;\;#\,\,$      &        
\vrule \hfil$\;\,#\,\,$      &        
\vrule \hfil$\;#\;$\hfil              
\vrule \\ 
\hline
\vphantom{$\sqrt A$}
 ZEF&\alpha^\circ\quad\;&\delta^\circ&N^*\;\,&\bar{N}\quad&\sigma\quad&D\\
\hline
 2 &        115\dots136 &  70\dots76 & 4724.2 & 4558.6 & 52.77 &  3.14  \\
\hline
 3 &        109\dots120 &  61\dots66 & 3580.6 & 3401.5 & 54.99 &  3.26  \\
   &        111\dots120 &  62\dots66 & 2315.8 & 2184.0 & 42.52 &  3.10  \\
Ir &        105\dots117 &  60\dots66 & 4765.7 & 4536.3 & 73.65 &  3.11  \\
\hline      
 9 (16) &   3\dots 19 &  47\dots57 &12266.9 &11977.8 & 93.13 &  3.10  \\
          &   6\dots 20 &  47\dots56 & 9744.3 & 9458.4 & 85.12 &  3.36  \\ 
          &   7\dots 21 &  46\dots55 & 9748.3 & 9478.5 & 74.96 &  3.60  \\ 
          &  10\dots 21 &  48\dots55 & 6027.6 & 5820.7 & 63.22 &  3.27  \\
          &  15\dots 21 &  47\dots51 & 1922.1 & 1808.1 & 33.86 &  3.37  \\ 
\hline
14 (14) & 175\dots180 &  19\dots24 & 379.5 & 327.1 & 14.59 &  3.60    \\
          & 175\dots197 &  20\dots39 &14014.1 &13689.8 &102.80 &3.15    \\
          & 176\dots190 &  26\dots38 & 6770.4 & 6564.0 & 64.96 &3.18    \\
          & 181\dots184 &  26\dots29 & 285.9 & 236.5 & 14.38 &  3.43    \\
\hline
16 (6)  &  12\dots 23 &  28\dots37 & 4136.0 & 3999.7 & 41.95 &  3.25  \\
          &  13\dots 26 &  23\dots34 & 4342.6 & 4203.5 & 43.34 &  3.21  \\
          &  19\dots 24 &  27\dots31 &  677.1 &  604.7 & 22.14 &  3.27  \\
          &  20\dots 23 &  28\dots31 &  331.5 &  283.0 & 14.22 &  3.42  \\  
\hline
30 (19) & 161\dots167 &   1\dots 7 &  22.0 &  10.2 &  3.01 &  3.90    \\ 
          & 162\dots183 &  -5\dots16 & 529.5 & 472.4 & 18.32 &  3.11    \\ 
          & 163\dots168 &   2\dots 7 &  16.9 &   7.9 &  2.51 &  3.58    \\ 
          & 175\dots183 &   5\dots13 & 107.2 &  78.4 &  8.59 &  3.35    \\ 
          & 176\dots179 &  10\dots13 &  32.6 &  18.4 &  4.24 &  3.36    \\ 
\hline
31 (4) &  334\dots339 &  11\dots16 & 112.7 &  81.8 &  9.05 &  3.42    \\ 
         &  336\dots339 &  11\dots14 &  40.7 &  22.9 &  4.54 &  3.93    \\ 
         &  336\dots339 &  13\dots16 &  51.5 &  35.2 &  5.08 &  3.21    \\ 
\hline
32 (68) & 242\dots258 & -13\dots 3 &  15.1 &   7.4 &  2.38 &  3.26    \\ 
          & 247\dots262 &  -9\dots 6 &  34.9 &  20.6 &  4.41 &  3.24    \\ 
          & 248\dots262 &  -8\dots 6 &  34.0 &  19.4 &  4.15 &  3.52    \\ 
          & 250\dots262 &  -6\dots 6 &  29.6 &  16.4 &  3.90 &  3.39    \\ 
          & 252\dots269 & -10\dots 7 &  49.0 &  32.8 &  4.97 &  3.26    \\ 
          & 265\dots272 &   4\dots11 &  58.9 &  38.5 &  5.92 &  3.45    \\ 
          & 268\dots272 &   7\dots11 &  29.7 &  16.9 &  3.74 &  3.41    \\ 
\hline
33 &        129\dots132 &   7\dots10 &  17.2 &   8.2 &  2.69 &  3.36    \\ 
Ir &        129\dots133 &   5\dots10 &  25.7 &  14.7 &  3.54 &  3.11    \\ 
\hline
}

\table{tab:PSRnearby}{
   Some parameters of the ZEF with pulsars at angular distances
   $0<\Delta\le3^\circ$.
   See \Tab{tab:SNRcells} for notation and parameters of ZEF No.~35.}{
\vrule \hfil$\,\,$#          &        
\vrule \hfil$\;\,#\,$        &        
\vrule \hfil$\;\,#\,\,$\hfil &        
\vrule \hfil$\;\,#\,\,$      &        
\vrule \hfil$\;\;#\,\,$      &        
\vrule \hfil$\;\,#\,\,$      &        
\vrule \hfil$\;#\;$\hfil              
\vrule \\ 
\hline
\vphantom{$\sqrt A$}
 ZEF&\alpha^\circ\quad\;&\delta^\circ&N^*\;\,&\bar{N}\quad&\sigma\quad&D\\
\hline
 4 (Ir) & 142\dots145 &  61\dots65 &  846.2 &  757.3 & 25.57 &  3.47 \\
\hline
 17     & 112\dots116 &  34\dots37 &  658.1 &  590.5 & 21.76 &  3.11 \\
\hline
21 (Ir) & 142\dots145 &  16\dots27 &  508.8 &  448.9 & 18.35 &  3.26 \\
\hline
 24     & 110\dots115 &  19\dots24 &  374.6 &  327.1 & 14.59 &  3.26 \\
        & 112\dots115 &  19\dots22 &  130.4 &  100.6 &  8.75 &  3.41 \\
\hline
26 (19) &  54\dots 59 &   5\dots10 &   32.5 &   18.4 &  4.46 &  3.17 \\
        &  54\dots 64 &   7\dots17 &  310.8 &  262.7 & 15.02 &  3.20 \\
        &  56\dots 62 &  11\dots17 &  173.2 &  132.2 & 11.80 &  3.47 \\
        &  57\dots 61 &  16\dots20 &  153.5 &  121.3 & 10.11 &  3.18 \\
        &  58\dots 62 &  13\dots17 &  100.5 &   69.7 &  8.28 &  3.72 \\
\hline
 27     & 192\dots195 &  17\dots20 &  102.1 &   73.9 &  8.19 &  3.44 \\
 Ir     & 192\dots196 &  15\dots20 &  179.2 &  140.5 & 11.26 &  3.44 \\
 Ir     & 191\dots195 &  15\dots20 &  177.3 &  140.5 & 11.66 &  3.15 \\
\hline
 29     & 112\dots115 &  13\dots16 &   52.5 &   35.2 &  5.04 &  3.43 \\
 Ir     & 107\dots116 &  12\dots16 &  160.1 &  129.4 &  9.36 &  3.28 \\
\hline
 34     &  82\dots 85 &   6\dots 9 &   13.8 &    6.3 &  2.30 &  3.28 \\
\hline
36 (2) & 69\dots 75 &   2\dots 8 &   25.1 &   14.0 &  3.47 &  3.20 \\
Ir       & 70\dots 75 &  -1\dots 8 &   23.6 &   12.9 &  3.41 &  3.15 \\
\hline
}

\subsection{Pulsars.}
   \Figure{fig:PSRs} shows the mutual position of the ZEF and pulsars
   known in the region $\delta\ge-15^\circ$~\cite{ATNF}, version of
   October~11, 2004.
   Forty-six of these 640 pulsars lie inside the ZEF, other 34 pulsars
   lie at angular distances $\Delta\le2^\circ$ from the boundaries of
   the nearest ZEF.
   As one can see from the figure, the list of zones with `neighbouring' 
   pulsars consists of 21 ZEF: No.~2--6, 8, 9, 14, 16, 18, 20, 21, 26,
   and 29--36.
   Notice that all ZEF discussed above in connection with galactic SNRs
   belong to the list.
   Some parameters of the ZEF `selected' by pulsars are presented in
   Tables~\tableref{tab:PSRinside} and~\tableref{tab:PSRnearby}.
   The biggest number of pulsars within $2^\circ$-neighbourhoods are
   found for ZEF No.~32 (26 PSRs), No.~8 and~18 (8 PSRs),
   No.~6 and No.~20 (7 and~6 PSRs respectively).
   Each of ZEF No.~5, 9, 14, and~21 has four neighbouring pulsars.
   It is interesting to mention that~13 of~640 pulsars in the region
   $\delta\ge15^\circ$ are associated with galactic SNRs.
   Five of them lie within $\Delta\le2^\circ$ from the ZEF.

   We have already mentioned the pulsar PSR B0656+14 in connection with
   the single-source model by Erlykin and Wolfendale~\cite{EW-JPG}.
   The pulsar has equatorial coordinates $\alpha=104.95^\circ$,
   $\delta=14.24^\circ$ and is located close in projection to the 
   center of the Monogem Ring SNR~\cite{Thorsett}.
   As is clear from \Fig{fig:PSRs} and \Tab{tab:PSRnearby}, the pulsar
   lies at the angular distance $\Delta=2.05^\circ$ from the left
   boundary of the outer cell of ZEF No.~29.
   This is an irregular cell of the size $9^\circ\times4^\circ$.
   The interior (regular) $3^\circ\times3^\circ$ cell is located at
   $\Delta=7.05^\circ$ from the pulsar.

   The interior cell is especially interesting in connection with the
   article by Chilingarian et al.~\cite{Chili03}, in which
   arrival directions of more than $2\cdot10^6$ EAS with the size
   $\Ne>10^5$ detected by the MAKET-ANI experiment have been analysed.
   As a result, a $3^\circ\times3^\circ$ CEF located at
   $\alpha=111^\circ\dots114^\circ$, $\delta=12.5^\circ\dots15.5^\circ$
   has been found.
   Evidently, this CEF is very close to the interior cell in ZEF No.~29.
   Here one should take into account that (i)~we have not studied
   cells with boundaries located at non-integer values of~$\alpha$
   or~$\delta$, and (ii)~Chilingarian et al.\
   have not studied cells with the size larger than $3^\circ\times3^\circ$.
   Thus we think this is a remarkable coincidence of the results,
   obtained in different experiments and by different methods of data
   analysis.

   It is worth mentioning that the Monogem Ring remnant covers a huge
   region with the diameter of approximately~$25^\circ$ with its most
   `pronounced' part being at the right hand side of the circle, see the
   figure in~\cite{Thorsett}.
   (To the contrary to our figures, Thorsett et al.~\cite{Thorsett} use
   the grid with  bigger values of~$\alpha$ located at the left-hand
   side of the figure.)
   Thus it is clear that ZEF No.~24 and~35 are intersected by the
   most pronounced part of the Monogem Ring, cf.\ \Fig{fig:SNRs}.
   
   As one can see in \Fig{fig:PSRs}, two other pulsars are located
   comparatively close to ZEF No.~29.
   These are J0711+0931, located on the top of ZEF No.~35,
   ($d=2.39$~kpc, $\Delta=2.5^\circ$ from ZEF No.~29) and
   J0751+1807 ($d=2.02$~kpc, $\Delta=2.7^\circ$), which lies to the NE
   from ZEF No.~29.
   (This pulsar looks to be located closer to ZEF No.~24 but in fact
   the corresponding angular distance equals nearly~$2.8^\circ$.)

   Another interesting issue is the distribution  of distances~$d$ to
   the `neighbouring' pulsars.
   As we have already mentioned above, there are 80 pulsars that
   belong to the $2^\circ$-neighbourhoods of the ZEF.
   One of them lies at $d>10$~kpc. 
   Distances to four other pulsars are unknown.  
   The remaining 75 pulsars are located at $d\le10$~kpc.
   As for the whole set of 640 pulsars in the region $\delta\ge-15^\circ$,
   582 of them have $d\le10$~kpc, and $d>10$~kpc for 49 pulsars.
   Distances to the remaining 9 pulsars are unknown.
   Thus let us take a look at the distribution of distances $\le10$~kpc.
   To do this, define
\[
   K(d) = \frac{582}{75} 
          \frac{N_{\rm PSR}^{\rm close}(d)}{N_{\rm PSR}^{\rm all}(d)},
   \label{eq:K}
\]
   where $N_{\rm PSR}^{\rm close}$ and $N_{\rm PSR}^{\rm all}$ are
   the number of pulsars within the corresponding intervals of~$d$
   for the $2^\circ$-neighbourhoods of ZEF and for
   all pulsars respectively.
   A value $K=1$ corresponds to the case when a fraction of 
   neighbouring pulsars at the given interval of~$d$ equals that for
   the whole set of pulsars.

\figure{fig:K}{
   The behaviour of the function $K(d)$, see Eq.~(\ref{eq:K}), for 0.5-kpc 
   intervals.}{\fig{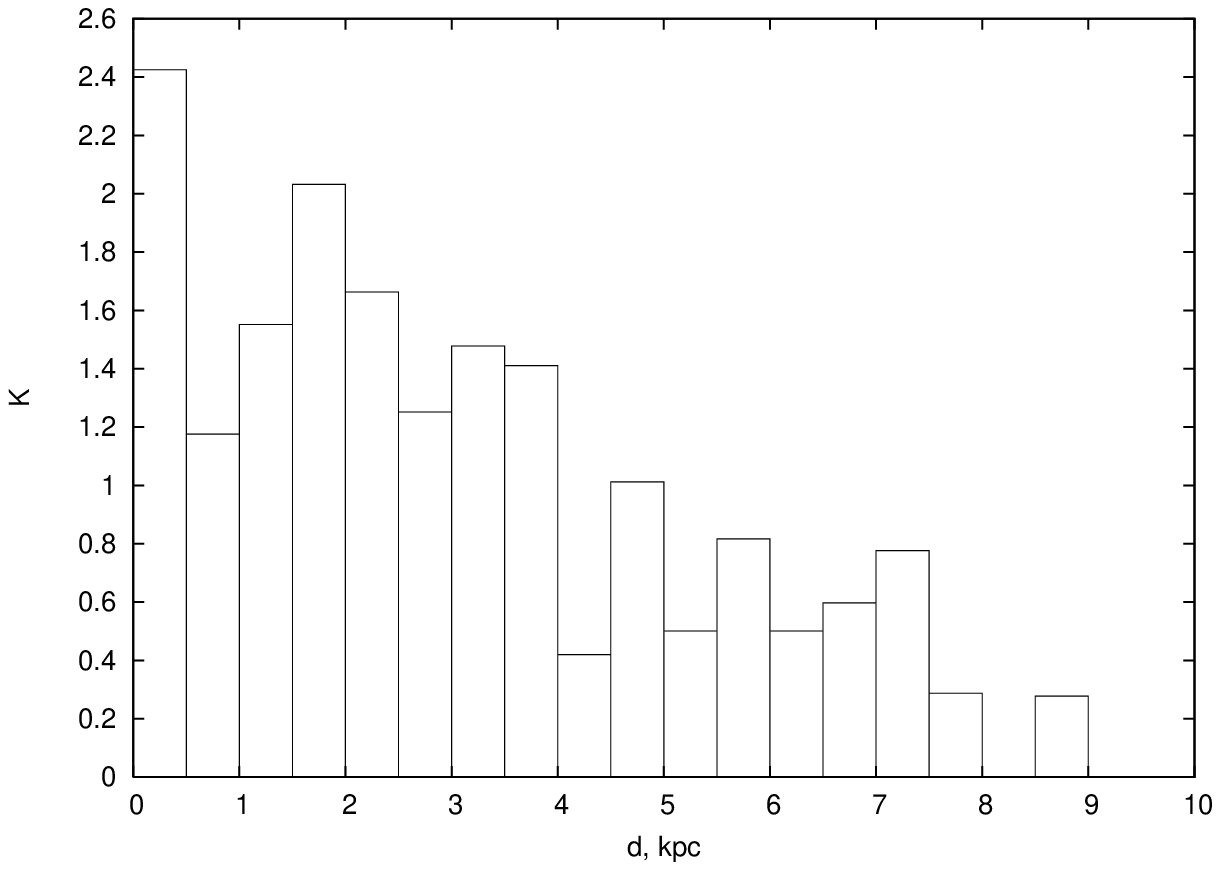}}
   
   \Figure{fig:K} shows the behavior of~$K(d)$ in 0.5-kpc intervals.
   It is clearly seen that a fraction of neighbouring pulsars
   with $d>8$~kpc is negligibly small.
   On the other hand, $K>1$ for $d\le4$~kpc with the most pronounced
   bins located at $1.5<d\le2$~kpc and $d\le0.5$~kpc.%
\footnote{%
   The ATNF database~\cite{ATNF} gives two `basic' values of
   distance to a pulsar: the best estimate of the pulsar distance and the
   distance based on the Taylor \& Cordes electron density
   model~\cite{TaylorCordes}. 
   For definiteness, we use the best estimate.
   Nevertheless, our results do not change qualitatively if the second
   value is used.  In this case, the highest bins correspond to distances
   $d\le0.5$~kpc and $1.5<d\le2.5$~kpc.
}   
   In our opinion, the latter fact can witness in favour of the hypothesis
   that a noticeable fraction of the overall flux of CRs near the knee
   is produced by nearby pulsars.

   Totally, we find five pulsars located at distances $d\le0.5$~kpc 
   from the solar system that belong to the $2^\circ$-neighbourhoods of ZEF:
\bitem
   PSR B0656+14 at $\Delta\approx2^\circ$ from ZEF No.~29 ($d=0.29$~kpc);
\bitem
   PSR J1136+1551 inside ZEF No.~30 ($d=0.36$~kpc),  also
   at $\Delta=1.5^\circ$ from ZEF No.~14;
\bitem   
   PSR J1744$-$1134 at $\Delta=1.6^\circ$ from ZEF No.~32 
   ($d=0.36$~kpc);
\bitem
   PSR J0814+7429 inside ZEF No.~2 ($d=0.43$~kpc);
\bitem
   PSR J0700+6418 inside ZEF No.~3 ($d=0.48$~kpc).
   
   The Geminga pulsar, which is often discussed as a possible source
   of PeV CRs ($d=0.16$~kpc) is located at the angular distance
   $\Delta=3.9^\circ$ from the bottom-right corner of ZEF No.~18.
   
\remark
   We do not really need irregular CEF for zones No.~4 and~33 to
   obtain 80 pulsars in the $2^\circ$-neighbourhoods of the ZEF.
   If we exclude them from consideration, the corresponding pulsars
   `move' from inside the ZEF to $\Delta=1.7^\circ$ and 
   $\Delta=0.8^\circ$ for ZEF No.~4 and~33 respectively.

   As we have stated above, there are 21 ZEF that have pulsars within
   $2^\circ$-neighbourhoods.
   Three more zones (No.~17, 24, and~27, cf.\ \Fig{fig:SNRs}) are added
   to the list if  we select pulsars located at $\Delta\le3^\circ$.
   (In this case, we find 23 more neighbouring pulsars.)
   An extension of the selection region up to $\Delta\le4^\circ$
   does not change the list of `selected' ZEF.
   Hence, there remain 13 ZEF without neighbouring pulsars and/or
   galactic SNRs.
   It is not a simple task to estimate a possible excess of the flux of
   CRs due to a contribution from a pulsar, see, e.g.,~\cite{EW04}, but 
   the number of neighbouring pulsars for some of the `selected' ZEF
   does not look to be `sufficient' either.
   In particular, this applies to huge zones No.~14, 16, and 30 with
   the first two of them lying in the region with a comparatively big
   number of EAS per unit cell, see \Tab{tab:PSRinside}.
   Thus we are lead to the necessity to analyse positions of other
   possible sources of PeV CRs.

\figure{fig:OSCs}{
   The ZEF and open star clusters~($\circ$) in the region $\delta\ge-15^\circ$.
   Filled circles denote clusters lying within $2^\circ$-neighbourhoods of 
   the ZEF.}{\bfig{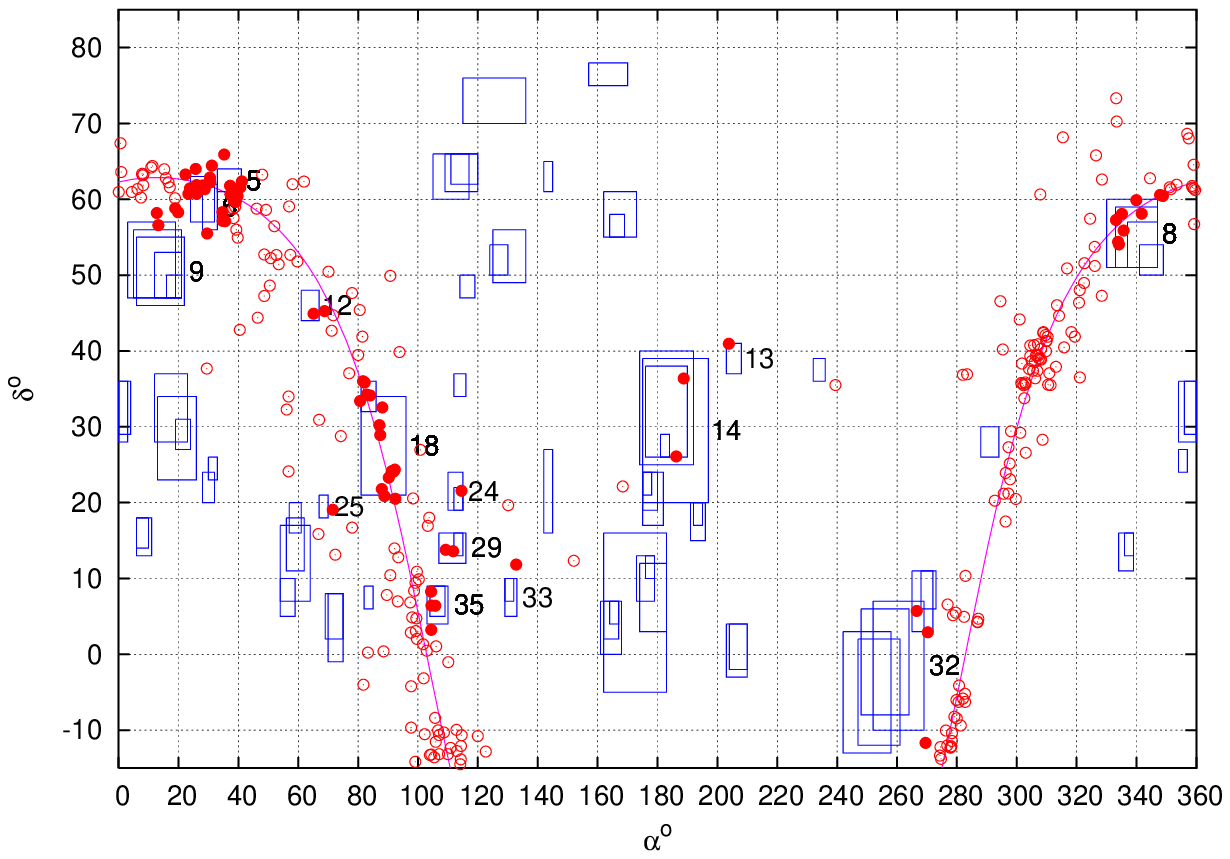}}

\subsection{Open Star Clusters and OB-associations.}
   The fact that most open star clusters (OSCs) are dominated by the hottest, 
   i.e., O-type and B-type, giant blue stars made us take a look at positions
   of these objects.
   We have analysed coordinates of 299 OSCs that lie in the region
   $\delta\ge-15^\circ$ and seem to be located in the Milky Way Galaxy.
   The list of these OSCs was mainly prepared with the use of
   the SIMBAD database~\cite{Simbad}.
   The result is shown in \Fig{fig:OSCs}: 42 OSCs lie inside the ZEF,
   32 more clusters are located at angular distances $0<\Delta\le2^\circ$
   from the ZEF.
   
   Similar to the cases of pulsars and galactic SNRs discussed above,
   ZEF No.~5, 6, 8, and~18 are among the `leaders' with the biggest
   number of objects lying within $2^\circ$-neighbourhoods.
   What seems to be more interesting is that the list of `selected'
   ZEF is now enriched with ZEF No.~12 (OSCs C~0417+448 and NGC~1605),
   ZEF No.~13 (OSC Czernik 35), and ZEF No.~25 (NGC~1647).
   In addition, ZEF No.~24, which has the J0751+1807 pulsar at the
   angular distance $\Delta=2.8^\circ$, now gets an object inside
   (NGC~2420).
   ZEF No.~15 is added to the list if we extend the selection region
   up to $\Delta=3^\circ$, cf.\ \Fig{fig:SNRs}.
   In this case, ZEF No.~25 and~26 also obtain a neighbouring OSC,
   namely the Hyades (C~0424+157).  
   It is worth mentioning that the Hyades is a nearby OSC, located
   at the distance of about $d=46$~pc from the solar system.
   
   It is interesting that ZEF No.~24, 29, and~35, discussed above in
   connection with the Monogem Ring SNR, contain 1, 2, and~3 OSCs 
   respectively {\it inside}.
   A small cell at the NW corner of ZEF No.~18 also becomes `selected'
   for the first time.
   In particular, it contains NGC~1931, M36, and M38 inside and
   M37 just nearby.
   (These three Messier objects are located at $d=1.26$--1.35~kpc.)
   Finally, let us mention an OSC that lies near the center of
   ZEF No.~14.
   This is the Coma star cluster, located at $d=90\pm5$~pc from the solar
   system.

\table{tab:OSCcells}{
   Parameters of ZEF `selected' by OSCs.
   See \Tab{tab:SNRcells} for notation.}{
\vrule \hfil$\,\,$#          &        
\vrule \hfil$\;\,#\,$        &        
\vrule \hfil$\;\,#\,\,$\hfil &        
\vrule \hfil$\;\,#\,\,$      &        
\vrule \hfil$\;\;#\,\,$      &        
\vrule \hfil$\;\,#\,\,$      &        
\vrule \hfil$\;#\;$\hfil              
\vrule \\ 
\hline
\vphantom{$\sqrt A$}
 ZEF&\alpha^\circ\quad\;&\delta^\circ&N^*\;\,&\bar{N}\quad&\sigma\quad&D\\
\hline
12 &  61\dots 67 &  44\dots48 & 1857.9 & 1751.9 & 33.22 &  3.19 \\
\hline
13 & 203\dots208 &  37\dots41 & 1280.9 & 1182.3 & 29.07 &  3.39 \\
\hline
15 & 232\dots236 &  36\dots39 &  731.9 &  657.9 & 23.60 &  3.14 \\
\hline
25 &  67\dots 70 &  18\dots21 &  115.1 &   86.6 &  8.79 &  3.24 \\
\hline
}

   Thus an assumption that open star clusters can be sources of PeV
   CRs adds three or four more ZEF to the list of zones with
   neighbouring CR sources.
   Their parameters are listed in \Tab{tab:OSCcells}.


   The so-called superbubble model~\cite{Parizot} stimulated us to analyse
   positions of OB-associations.
   We have found that at least seven OB-associations are located at
   angular distances $\Delta\le2^\circ$ from ZEF No.~6, 8, and~36.
   Three other associations are located at 
   $\Delta\approx2.2^\circ$--$2.5^\circ$ from ZEF No.~20.
   Remarkably, open star clusters and OB-associations taken together
   with galactic SNRs `select' all ZEF in the vicinity of the galactic plane.


\figure{fig:galH2}{
   The ZEF and galactic HII regions~($+$)~\cite{galH2}.
   Filled circles show regions lying within $2^\circ$-neighbourhoods of 
   the ZEF.}{\bfig{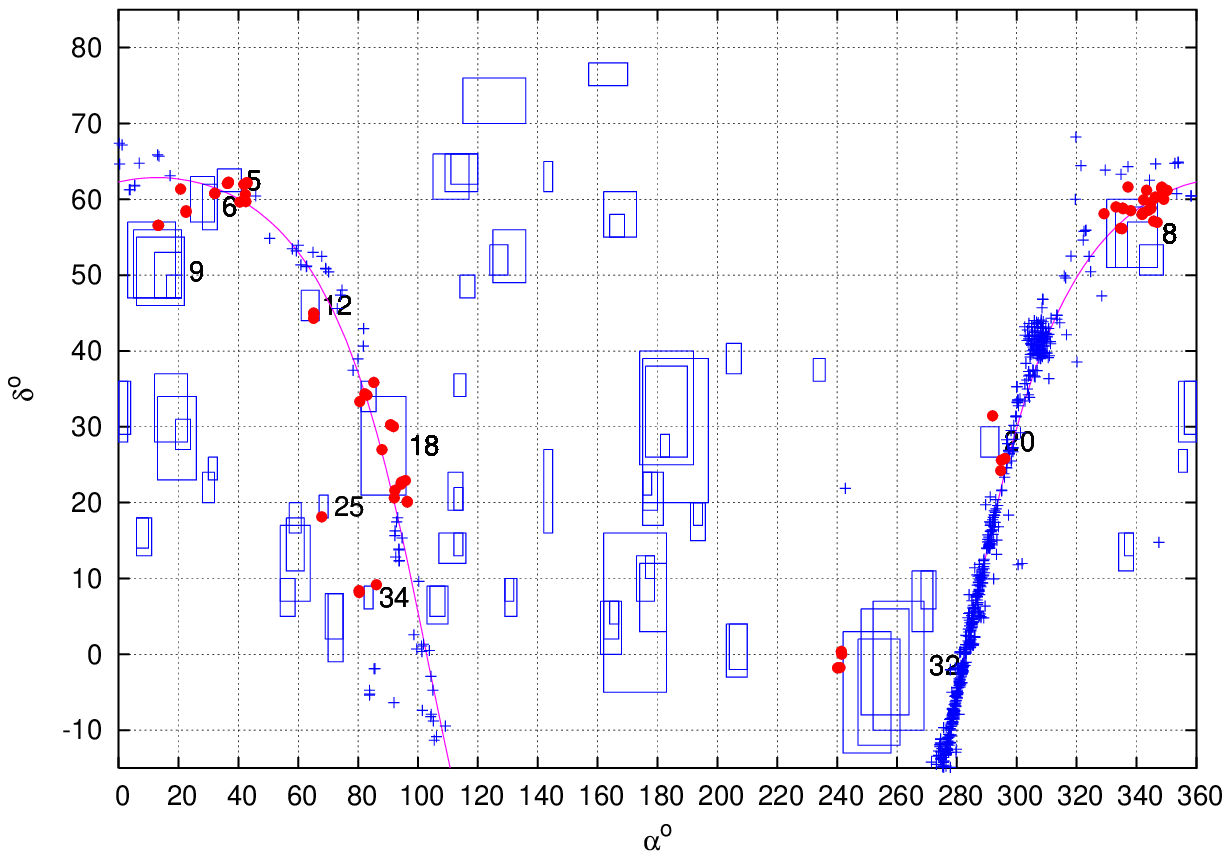}}

\subsection{Galactic HII Regions.}
   Another group of astrophysical objects that may be considered as
   possible sources/accelerators of cosmic rays consists of regions of
   ionized hydrogen~(HII).
   We have analysed positions of galactic HII regions given in the
   catalogue by Paladini et al.~\cite{galH2}.
   The result is shown in \Fig{fig:galH2}.

   Seventy-three of 728 HII regions with declination $\delta\ge-15^\circ$
   belong to the $2^\circ$-neighbourhoods of the ZEF.
   Thirty-seven of these regions lie inside the corresponding ZEF.
   The biggest number of HII regions is found in the vicinity
   of ZEF No.~8 (29 regions, 17 of them inside the zone),
   ZEF No.~18 (15 and 11 regions respectively), and ZEF No.~5
   (9 and 2 regions respectively).
   Similar to the case of open star clusters, ZEF No.~12 and~25,
   which do not have any galactic SNRs or pulsars in their vicinity,
   become `selected' with respectively three and one HII regions 
   inside.
   An additional `support' is also provided to ZEF No.~34
   (3 HII regions nearby) and to the small cell in the NW corner
   of ZEF No.~18.
   Still, the analysis of positions of galactic HII regions
   has not added new zones to the list of ZEF with possible
   sources and/or accelerators of PeV CRs in their vicinity.
   
   An analysis of positions of Wolf--Rayet (WR) stars did not reveal
   anything remarkable except that there are more than 100 WR stars
   within the boundaries of a small  $5^\circ\times4^\circ$ cell located
   inside ZEF No.~16.
   We shall discuss this fact later.
   At the moment, we see that 25 of 37 ZEF have at least one galactic 
   object of the types discussed above in their $2^\circ$-neighbourhoods,
   28 ZEF if we consider $3^\circ$-neighbourhoods.
   Thus, there remain at least 9 ZEF (No.~1, 7, 10, 11, 19, 22, 23, 28,
   and~37) that do not have known galactic objects of the above types in
   their vicinity.
   A number of the zones may look `underfilled'.
   If we believe that presented ZEF somehow reflect a situation with the
   angular distribution of the flux of PeV cosmic rays then the
   appearance of `empty' and `underfilled' ZEF raises a question on
   other possible CR sources.
   An unexpected hint comes from an analysis of arrival directions
   of the ultra-high energy cosmic rays (UHECRs) registered with the 
   AGASA array.

\figure{fig:AGASA}{
   The ZEF and arrival directions of the UHECRs with 
   $E\gtrsim4\cdot10^{19}$~eV registered with the AGASA 
   array~($+$)~\cite{AGASA}.
   Filled circles mark UHECRs lying within $4^\circ$-neighbourhoods of 
   the ZEF.}{\bfig{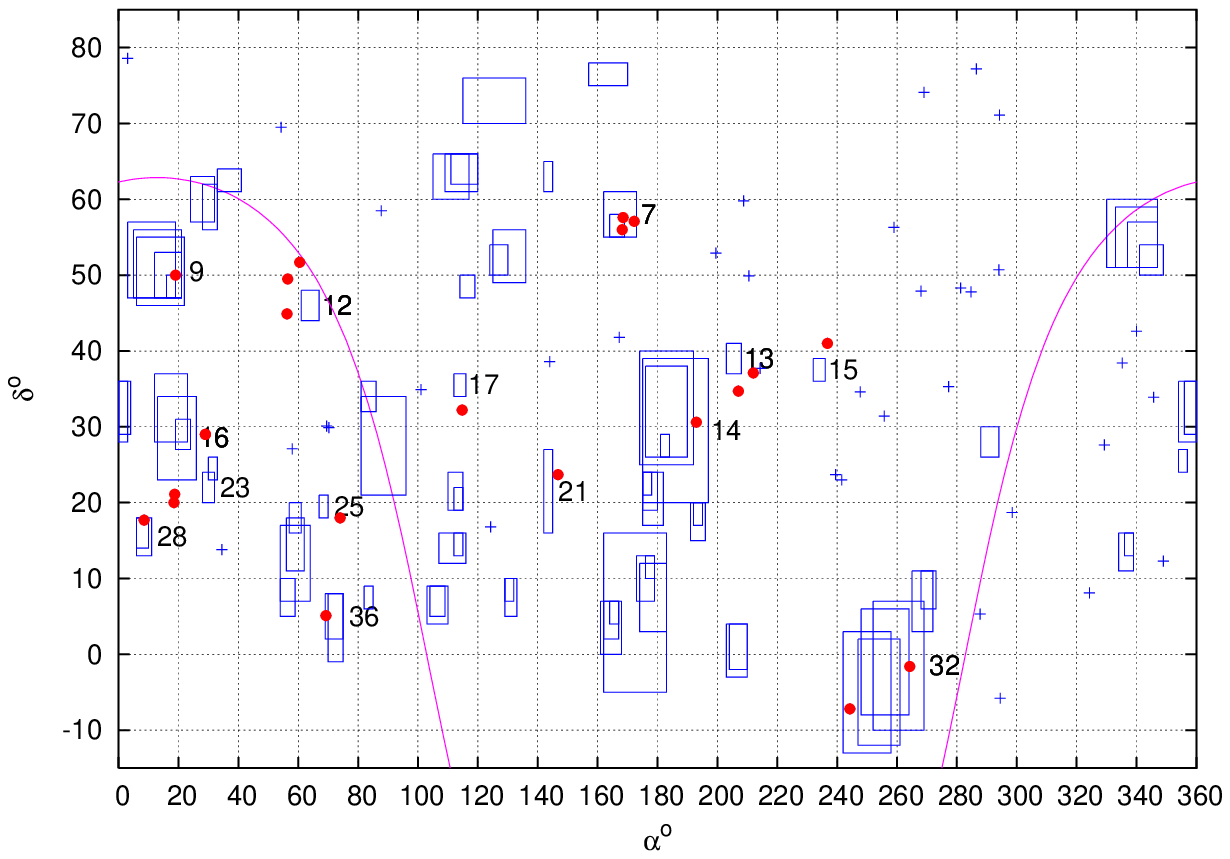}}

   Earlier, we have used the same data set for a study of arrival times of 
   EAS.
   In particular, we found 20 groups (`cluster events') of consecutive EAS 
   that produced bursts of the count rate~\cite{IzvRAN01}, \cite{preprint}.
   The majority of EAS within these groups did not have close arrival
   directions but certain clusters  covered approximately the same
   regions of the sky.
   Surprisingly, we have found a group of four such clusters with the 
   famous AGASA `C2' triplet near the center of the corresponding 
   region, see Fig.~30 in~\cite{preprint}.
   This gave rise to an idea that there may be some connection between
   UHECRs and cosmic rays in the PeV range.
   The present investigation gave us a chance to test the idea.
   
   \Figure{fig:AGASA} shows arrival direction of 58 UHECRs with energies
   $E\gtrsim4\cdot10^{19}$~eV registered with the AGASA array~\cite{AGASA}.
   Twenty-one of these UHECRs belong to $4^\circ$-neighbourhoods of the ZEF.
   It is remarkable that a number of ZEF that have very few or no
   neighbouring pulsars and galactic SNRs become `selected' by the
   AGASA UHECRs (namely, ZEF No.~7, 12--17, 23, 25, 28, 
   and~36) with the C2 triplet lying {\it exactly inside\/} ZEF No.~7.%
\footnote{%
   A recent report by Abbasi et al.~\cite{HiRes} has revealed that
   one of the UHECRs with $E>3.0\times10^{19}$~eV recorded by the HiRes
   experiment is located at $\alpha=169.0^\circ$ and $\delta=55.9^\circ$
   and thus also lies inside ZEF No.~7.
}
   This observation motivated us to compare coordinates of the ZEF with
   positions of the nearby active galactic nuclei and interacting
   galaxies (IGs).

\figure{fig:AGNIGs}{
   The ZEF, AGN ($\triangle$), and interacting galaxies~($\circ$)
   at redshift $z\le0.01$~\cite{Simbad}.
   Filled triangles and filled circles mark respectively
   AGN and IGs that lie within $4^\circ$-neighbourhoods of the ZEF.
   The $\cap$-like curve shows the Supergalactic plane.}{\bfig{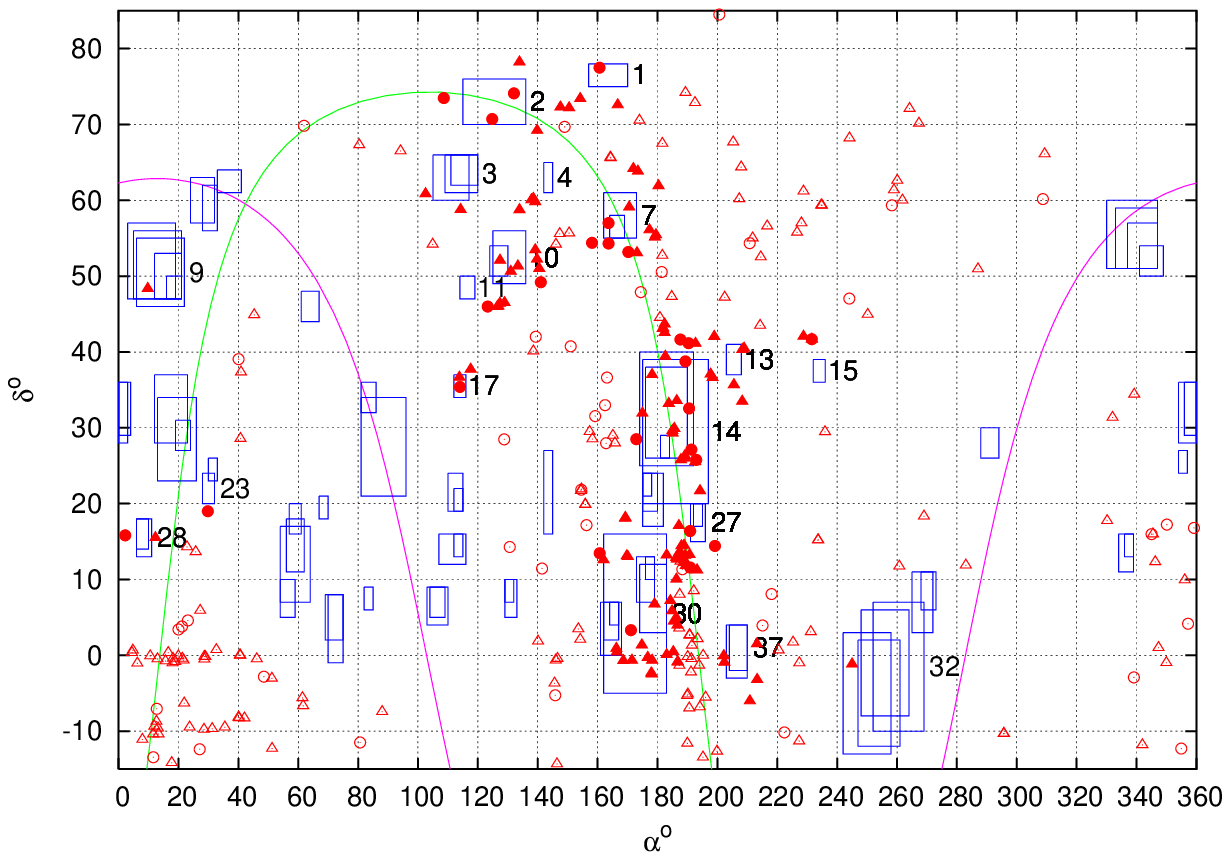}}

\subsection{Interacting Galaxies and Active Galactic Nuclei.}
   It has been demonstrated in~\cite{Tarle} that a number of particles with
   energies $E<10^6$~GeV/nucleon that have had time to arrive to our galaxy
   from distances greater than $\sim100$~Mpc is completely negligible.
   Thus we restrict our analysis to the sources located at redshift
   $z\le0.01$, which corresponds to $d\approx40$~Mpc assuming the
   Hubble constant $H=75$~km$\,$sec$^{-1}\,$Mpc$^{-1}$.
   As it will be demonstrated below, in many cases we need far less values
   of~$z$ to find an AGN near a ZEF.

\table{tab:AGN}{
   A list of some AGN and interacting galaxies (IGs) located at
   redshifts $z\le0.01$ and lying within $4^\circ$-neighbourhoods of
   the ZEF. 
   Types of the AGN:
   `Sy' is for Seyfert galaxies, possibly with their type (1 or~2), 
   `LINER' is for Low-Ionization Nuclear Emission-line Regions, 
   and `BLL' is for a BL Lac object.
   For ZEF with nearby AGN, the number of nearby AGN at redshifts $z\le0.005$
   and $z\le0.01$ respectively are given in parenthesis.
   Names of AGN beginning with~`J' correspond to the Veron-Cetty~\& Veron
   catalogue~\cite{VV2003c}.
   For IGs, we give nomenclature used in the SIMBAD database~\cite{Simbad}
   with data given according to~\cite{APG}.
}{
\vrule \hfil$\,\,$#          &        
\vrule \hfil$\,\,#\,\,$\hfil &        
\vrule \hfil$\,\,#\,\,$\hfil &        
\vrule \hfil$\,\,#\,\,$\hfil &        
\vrule $\,\,$#\hfil          &        
\vrule $\;\;$#$\;\,$\hfil    &        
\vrule \hfil$\,\,#\,\,$               
\vrule \\ 
\hline
ZEF\quad$\;$&\alpha^\circ&\delta^\circ&\Delta^\circ&Name&Type&    z\qquad \\
\hline
1 (1/4)  
   & 147.59 & 72.28 & 3.79 &  NGC 2985                  & Sy1&    0.004306\\
   & 160.66 & 77.49 & 0.00 &  APG 156                   & IG &    0.006274\\
\hline
2 (3/3)
   & 139.83 & 69.20 & 1.55 &  NGC 2787                  & LINER&  0.002298\\
   & 124.77 & 70.72 & 0.00 &  APG 268                   & IG &    0.000524\\
\hline
3 (0/2)
   & 114.24 & 58.77 & 1.23 &  Mrk 9                     & Sy1&    0.006321\\
   & 102.54 & 60.85 & 1.20 &  NGC 2273                  & Sy2&    0.006151\\
\hline
4 (2/3)
   & 137.91 & 60.04 & 2.23 &  NGC 2768                  & Sy &    0.004590\\
\hline
7 (7/8)     
   & 173.14 & 53.07 & 1.93 &  NGC 3718                  & LINER&  0.003306\\
   & 170.57 & 59.07 & 0.00 &  NGC 3642                  & LINER&  0.005327\\
\hline
10 (6/11)
   & 133.39 & 51.31 & 0.00 &  NGC 2681                  & LINER&  0.002308\\
   & 133.90 & 58.73 & 2.73 &  NGC 2685                  & Sy2&    0.002915\\
\hline
11 
   & 123.31 & 45.99 & 3.13 &  APG 6                     & IG &    0.001491\\
\hline
13 (3/6)
   & 205.53 & 35.65 & 1.35 &  NGC 5273                  & Sy1&    0.003596\\
   & 198.96 & 42.03 & 3.20 &  M 63                      & LINER&  0.001678\\
\hline
14 (19/23)
   & 194.18 & 21.68 & 0.00 &  M 64                      & Sy &    0.001341\\
   & 185.03 & 29.28 & 0.00 &  NGC 4278                  & LINER&  0.002145\\
   & 186.45 & 33.55 & 0.00 &  NGC 4395                  & Sy1&    0.001061\\
   & 192.72 & 41.12 & 1.25 &  M 94                      & Sy &    0.001024\\
   & 189.34 & 38.75 & 0.00 &  APG 211                   & IG &    0.001484\\
   & 190.38 & 41.15 & 1.15 &  APG 23                    & IG &    0.001811\\
\hline
15 (0/2) 
   & 231.53 & 41.67 & 2.69 &  NGC 5929                  & LINER&  0.008912\\
   & 228.76 & 42.05 & 3.92 &  NGC 5899                  & LINER&  0.008503\\
\hline
17 
   & 113.99 & 35.38 & 0.00 &  APG 250                   & IG &    0.004543\\
\hline
23 
   &  29.83 & 19.01 & 0.99 &          NGC 772           & IG &    0.008199\\
\hline
27 (8/11)
   & 188.88 & 12.22 & 3.46 &  NGC 4550                  & LINER&  0.001338\\
   & 188.92 & 12.56 & 3.17 &  M 89                      & Sy2&    0.000914\\
   & 189.21 & 13.16 & 2.53 &  M 90                      & Sy &    -0.000721\\
   & 188.86 & 14.50 & 2.13 &  M 91                      & LINER&  0.001644\\
\hline
28 (0/1)
   &  12.27 & 15.49 & 1.22 &  J004903.7+152907          & Sy1&    0.010   \\
   &   2.19 & 15.82 & 3.66 &  APG 235                   & IG &    0.002885\\
\hline
30 (14/29)
   & 178.16 & -2.47 & 0.00 &  Mrk 1307                  & Sy1&    0.003426\\
   & 186.34 & 10.02 & 3.29 &  NGC 4380                  & BLL&    0.003229\\
   & 161.96 & 12.58 & 0.04 &  M 105                     & LINER&  0.002922\\
   & 186.27 & 12.89 & 3.18 &  M 84                      & Sy2&    0.003369\\
   & 170.06 & 12.99 & 0.00 &  M 66                      & LINER&  0.002425\\
   & 186.94 & 13.01 & 3.84 &  NGC 4438                  & LINER&  0.000234\\
   & 169.73 & 13.09 & 0.00 &  M 65                      & LINER&  0.002692\\
\hline
37 (0/5)
   & 213.31 & -3.21 & 3.31 &  NGC 5506                  & Sy2&    0.006068\\
   & 202.34 & -0.94 & 0.66 &  J132921.3$-$005639        & Sy1&    0.010   \\
\hline
}

\table{tab:AGNIGcells}{
   Parameters of ZEF `selected' by AGN and interacting galaxies
   at redshifts $z\le0.01$.
   See \Tab{tab:SNRcells} for notation.}{
\vrule \hfil$\,\,$#          &        
\vrule \hfil$\;\,#\,$        &        
\vrule \hfil$\;\,#\,\,$\hfil &        
\vrule \hfil$\;\,#\,\,$      &        
\vrule \hfil$\;\;#\,\,$      &        
\vrule \hfil$\;\,#\,\,$      &        
\vrule \hfil$\;#\;$\hfil              
\vrule \\ 
\hline
\vphantom{$\sqrt A$}
 ZEF&\alpha^\circ\quad\;&\delta^\circ&N^*\;\,&\bar{N}\quad&\sigma\quad&D\\
\hline
 1 & 157\dots170 &  75\dots78 & 1151.3 & 1036.5 & 35.80 &  3.21 \\
\hline
 7 & 162\dots173 &  55\dots61 & 4908.3 & 4688.6 & 69.51 &  3.16 \\
   & 164\dots169 &  55\dots58 & 1177.3 & 1089.2 & 27.86 &  3.16 \\
\hline
10 & 124\dots130 &  50\dots54 & 1945.7 & 1818.6 & 36.91 &  3.44 \\
   & 125\dots136 &  49\dots56 & 5984.6 & 5793.9 & 59.34 &  3.21 \\
\hline
11 & 114\dots119 &  47\dots50 & 1223.3 & 1127.1 & 30.18 &  3.19 \\
\hline
23 &  28\dots 32 &  20\dots24 &  263.0 &  223.6 & 11.95 &  3.30 \\
   &  30\dots 33 &  23\dots26 &  209.2 &  171.3 & 12.10 &  3.13 \\
\hline
28 &   6\dots 10 &  14\dots18 & 115.1 &  84.9 &  9.03 &  3.35 \\
   &   6\dots 11 &  13\dots18 & 159.1 & 121.2 & 10.48 &  3.62 \\
\hline
37 & 203\dots210 &  -3\dots 4 &  11.2 &   4.5 &  1.97 &  3.42 \\
   & 204\dots210 &  -2\dots 4 &  10.3 &   3.7 &  1.73 &  3.85 \\
\hline
}

   \Figure{fig:AGNIGs} presents positions of the AGN and interacting 
   galaxies at redshifts $z\le0.01$.
   Totally, 60 of 119 AGN that have redshifts $z\le0.005$ and are
   located in the region
   $\delta\ge-17^\circ$ belong to the $4^\circ$-neighbourhoods of the ZEF.
   Twenty of them lie inside the corresponding ZEF.
   The same numbers for 249 AGN at redshifts $z\le0.01$ are equal to
   106 and 32 objects respectively.
   As for 68 IGs, the corresponding numbers equal 26 and~10,
   see \Tab{tab:AGN} for a list of some of these objects.
   In its turn, \Tab{tab:AGNIGcells} presents parameters of ZEF
   `selected' by the AGN and interacting galaxies.

   Let us discuss these results in more details.
   First of all, as is clear from \Fig{fig:AGNIGs}, a number of
   the ZEF cover regions in the vicinity of the Supergalactic plane.
   ZEF No.~1, 7, 10, 11, 23, 28, and 37, which do not have neighbouring
   galactic sources of the types discussed above, become `selected'
   by the AGN and/or IGs.
   Further `support' is provided to ZEF No.~2, 3, 4, 13, 15, and~17.
   Perhaps, the most exciting observation is that there
   are numerous AGN and IGs within and near huge but `underfilled' zones 
   No.~14 and~30 with a hole bunch of galaxies located between ZEF
   No.~27 and No.~30.
   In sum, ZEF No.~14 has 23 neighbouring AGN and 8~IGs with respectively
   19 and~7 of them at redshifts $z<0.005$.
   Twenty-nine AGN and 2~IGs belong to the $4^\circ$-neighbourhood of
   ZEF No.~30.  
   Fourteen of these AGN and both IGs have $z<0.005$.
   Finally, all three IGs and 8~of 11 AGN that lie in the 
   $4^\circ$-neighbourhood of ZEF No.~27 are located at redshifts $z<0.005$
   (with the M58 galaxy at $z=0.005047$).
   What seems to be even more important is that almost all galaxies
   selected as being close to ZEF No.~27 belong to the Virgo cluster
   of galaxies.
   These are NGC~4762, NGC~4550, NGC~4639, NGC~4477, M58, M88,   
   M89, M90, M91 with the famous M87 galaxy located at $\Delta=4.1^\circ$
   from ZEF No.~27.
   Galaxies M84, NGC~4380, and NGC~4438, which also belong to the 
   Virgo cluster, lie closer to ZEF No.~30.
   Remarkably, NGC~4380 is a BL~Lac object.
   In addition, one can notice that the Coma cluster and Coma
   supercluster of galaxies lie {\it inside\/} ZEF No.~14.
   (Their coordinates are  $\alpha\approx195^\circ$, $\delta\approx28^\circ$
   and $\alpha\approx186^\circ$, $\delta\approx24^\circ$ 
   respectively~\cite{NED}.)
   In our opinion, all this makes a basis for a conjecture that zones
   No.~30, 27, 14, and possibly No.~13 may be considered as parts of one
   huge ZEF, which to big extent originates from a contribution to
   the overall CR flux from the Coma--Virgo cluster of galaxies.

   As we have already mentioned above, AGN and IGs provide additional
   `support' to comparatively large ZEF No.~2 and No.~3 `selected'
   by pulsars but `select' a large zone No.~37 for the first time.
   On the other hand, it is hard to believe that AGN and/or IGs or
   even their compact groups can produce small ZEF of CRs in the PeV
   region as those No.~4, 11, 13, 15, 17, 23, and~28.
   It seems more likely that these ZEF should originate from a
   contribution from galactic sources.
   Still, we have seen that ZEF No.~11, 23, and~28 do not have
   neighbouring objects of the types considered above.
   The same is true for ZEF No.~7.
   Besides this, one cannot help mentioning that ZEF No.~7, 13, 15, 
   17, and~28 are `selected' by both the UHECRs registered with the 
   AGASA array on the one hand and AGN and IGs (except for ZEF No.~13) 
   on the other.
   Thus we are lead to the following conjectures:
   (i)~the corresponding UHECRs originate from the AGN selected in
   the neighbourhoods of these ZEF,
   (ii)~a part of UHECRs are neutral particles or there are certain
   voids in intergalactic magnetic fields,
   (iii)~a part of ZEF, especially those far from the galactic plane,
   are formed with a contribution of extragalactic CRs.

   Notice also that another AGASA UHECR lies inside ZEF No.~14 and has
   a number of AGN around, cf.\ Figures~\figref{fig:AGASA}
   and~\figref{fig:AGNIGs}.
   On the other hand, three other AGASA UHECRs lie within ZEF No.~9
   and~32, which have numerous pulsars inside but just one AGN each,
   both located comparatively far from the arrival directions of the
   UHECRs.
   In our opinion, this fact may witness in favour of an idea that a
   part of UHECRs may originate from pulsars located in the Galaxy.
   This idea has already been discussed elsewhere, see, 
   e.g.,~\cite{AGASA99} and references therein.

\subsection{ZEF No.~16.}
   Finally, let us briefly discuss ZEF No.~16.
   Recall that this ZEF has just one distant pulsar at its left border,
   see \Fig{fig:PSRs}, and no other potential sources of CRs nearby
   but more than 100 WR stars located inside its small central cell.
   (Also notice that there is an AGASA UHECR located to the right of
   ZEF No.~16, see \Fig{fig:AGASA}.)
   Still, the zone lies in the Supergalactic plane.
   A key to a possible explanation of the appearance of ZEF No.~16
   comes from an observation that all these WR stars belong to the
   M33 galaxy, a prominent member of the Local Group of galaxies
   ($\alpha\approx23.46^\circ$, $\delta\approx30.66^\circ$;
   $d=730\pm168$~kpc~\cite{Brunthaler}).
   What seems to be really important is that M33 hosts the most luminous
   steady X-ray source in the Local Group, see~\cite{M33} and
   references  therein, a number of radio sources, and more than~80 HII
   regions~\cite{NED} including NGC~588, NGC~595, and NGC~604, a giant
   diffuse nebula, nearly 1500 light-years across. 
   Thus we expect that ZEF No.~16 may originate from a contribution from
   these objects.

   Interestingly, one of successive air shower events in the 50~TeV--10~PeV
   energy range reported by the LAAS group~\cite{LAAS} has
   the averaged `coordinates' $\alpha=21.75^\circ$, $\delta=30.81^\circ$
   and thus also lies within the central cell of ZEF No.~16.
   
\subsection{`Empty' ZEF.}
   We are thus left with two ZEF without any object of the types discussed 
   above in their vicinity.
   These are ZEF No.~19 and~22.
   There naturally appears a question of why are they `empty'?
   
\table{tab:empty}{%
   Parameters of `empty' ZEF No.~19 and~22.}{                             
\vrule \hfil$\,\,$#          &        
\vrule \hfil$\;\,#\,$        &        
\vrule \hfil$\;\,#\,\,$\hfil &        
\vrule \hfil$\;\,#\,\,$      &        
\vrule \hfil$\;\;#\,\,$      &        
\vrule \hfil$\;\,#\,\,$      &        
\vrule \hfil$\;#\;$\hfil              
\vrule \\ 
\hline
\vphantom{$\sqrt A$}
 ZEF&\alpha^\circ\quad\;&\delta^\circ&N^*\;\,&\bar{N}\quad&\sigma\;\;&D\\
\hline
 19 & 354\dots363 &  28\dots36 & 2927.9 & 2809.5 & 38.15 &  3.10 \\
    & 356\dots364 &  29\dots36 & 2377.0 & 2267.3 & 34.95 &  3.14 \\
\hline
 22 & 354\dots357 &  24\dots27 &  233.5 &  190.3 & 13.56 &  3.19 \\
\hline
}

   Both `empty' ZEF are made of regular cells. 
   (They join into one bigger cell if we consider irregular cells.)
   Parameters of these two ZEF are listed in \Tab{tab:empty}.
   One can see that there is nothing special, perhaps excluding the fact
   that the values of deviation~$D$ are close to the lowest allowed boundary.
   Still, this is not unusual for non-empty ZEF either, as 
   is clear from Tables~\tableref{tab:SNRcells}--\tableref{tab:OSCcells}
   and~\tableref{tab:AGNIGcells}.
   It is possibly more important that ZEF No.~19 intersects the
   boundary $\alpha=360^\circ$, and ZEF No.~22 lies close to it.
   At the moment, we cannot completely exclude a possibility that the
   appearance of ZEF No.~19 and~22 is an artifact of the algorithm
   resulting from a boundary effect.
   It is remarkable though that two of the eleven point sources in the
   100~GeV to 100~TeV energy range with an excess greater than~$4\sigma$,
   reported recently by the Milagro Collaboration~\cite{Milagro} are
   located inside ZEF No.~19.
   In addition, averaged coordinates of another successive air shower
   event reported by the LAAS group~\cite{LAAS} lie within
   the boundaries of ZEF No.~19.

\section{EAS with $\Ne\ge2\cdot10^5$}{sec:Ne2e5}
   It is undoubtedly interesting to compare arrival directions of CRs with
   energies just below the knee with those above it.
   Unfortunately, as we have already explained in Sec.~\sref{sec:data},
   the data set under consideration mostly covers the interval of
   energies just below the knee.
   Still, we find it interesting to perform a brief analysis of
   arrival directions of 142,223 EAS with $\theta\le45^\circ$ and
   $\Ne\ge2\cdot10^5$, which corresponds to energies 
   $E\gtrsim2\cdot10^{15}$~eV.

\figure{fig:Ne2e5}{
   Mutual positions of the ZEF obtained for the whole data set
   and for EAS with $\Ne\ge2\cdot10^5$
   (rectangles in thin and thick lines respectively.)}{\bfig{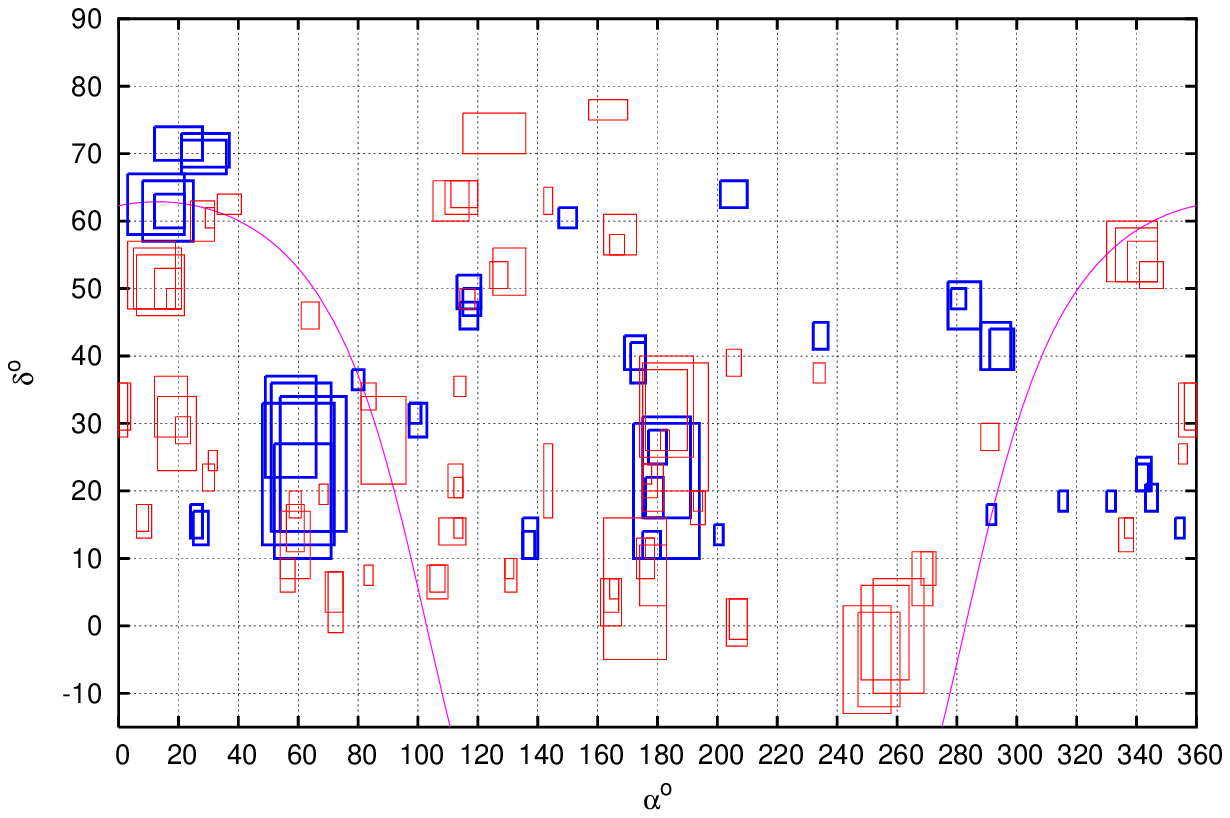}}

   After applying the same procedure of data analysis  (with the same
   random seeds used at the stage of the data alignment) we are left
   with 111,370 EAS and 21 new ZEF combined of 147 regular cells, see
   \Fig{fig:Ne2e5}.
   We do not think that $1.1\cdot10^5$ showers are sufficient to make
   definite conclusions on a distribution of their arrival directions
   and will thus only give brief comments.
   
   First of all, one can see that there is a huge new ZEF that partially 
   overlaps with `old' ZEF No.~14, 27, and~30.
   This new ZEF includes all AGN and IGs selected earlier that belong
   to the Virgo cluster of galaxies and those in the region of the Coma
   cluster.
   It seems to be quite natural since one may expect that AGN normally
   produce more energetic particles than pulsars and SNRs.
   Next, there is another huge ZEF which includes `old' ZEF No.~25 and
   a part of ZEF No.~26.
   This ZEF occurs to be mostly filled with OSCs and pulsars.
   One can also see a ZEF that is approximately adjacent to ZEF No.~6 and~9.
   Both SNRs located near the top left corner of ZEF No.~6
   (G126.2+1.6 and G127.1+0.5) belong to this new ZEF.
   Besides this, it also contains the famous Tycho SNR (SN1572).
   There are also a number of other new ZEF that lie close to the old
   ones.
   Finally, one can see a number of new ZEF that occupy areas of their
   own.
   All new ZEF have these or that astrophysical objects of the types
   discussed above in their vicinity.

\section{Conclusions}{sec:concs}
   We have presented the results of an analysis of arrival directions
   of more than $10^6$ EAS in the energy range $\approx0.1$--10~PeV
   registered with the EAS--1000 Prototype Array.
   An iterative algorithm developed for this investigation allowed us
   to find a number of zones with an excess of CR flux in the
   data set under consideration at $\ge3\sigma$ level.
   The zones are mostly located in the vicinity of the galactic and 
   Supergalactic planes.
   The majority of the ZEF have powerful galactic sources of CRs inside
	or nearby, namely, SNRs (with the Crab Nebula, Cassiopeia~A,
   and the Monogem Ring among them), pulsars, open star clusters,
	OB-associations, and regions of ionized hydrogen.
   This observation seems to witness in favour of the traditional
   point of view
   that objects of all these types make a contribution to the flux of
   PeV cosmic rays.

   On the other hand, there are a number of ZEF that have none or just
   a few objects of the above types in their vicinity.
   We find that one of such ZEF contains the M33 galaxy inside, while
	the majority of others have  neighbouring active galactic nuclei
	and/or interacting galaxies at redshifts $z<0.01$ with a big
	group of them located in the Virgo cluster of galaxies.
   These observations together with a number of coincidences between
   arrival directions of UHECRs with $E\gtrsim4\cdot10^{19}$~eV
	registered  with the AGASA array and positions of some ZEF lead us to
	the following conjectures:
   (i)~the flux of PeV cosmic rays is a superposition of two components,
   a galactic and an extragalactic ones (cf.~\cite{Muraishi}),
   and at least a part of the extragalactic component `remembers' its
   origin;
   (ii)~cosmic rays in the PeV (and possibly higher) energy range
   originate from astrophysical objects of {\it various\/} types;
   (iii)~there are astrophysical sources that emit/accelerate particles 
   in a wide range of energies so that a part of both PeV and EeV cosmic
   rays may have common origin.
   (This is not a new idea.  One can find a number of models which
   attempt to explain the origin of CRs of very different energies with
   the help of a unique type of sources and/or mechanism, see, 
   e.g., \cite{Parizot}, \cite{EWW01}, \cite{Wick}, and~\cite{Dar}.)
   It may be rather difficult though to explain how extragalactic sources
   can form a noticeable flux of PeV CRs from directions close to
   their origin.
   Perhaps, the most obvious ideas are that at least a part of the
   extragalactic component consists of neutral particles {\it or\/} that
   there are certain voids in intergalactic magnetic fields.

   The presented results pose a number of tasks for the future
   investigation.
   First of all, it is desirable to perform a comprehensive analysis
   of the algorithm with the use of a Monte-Carlo simulation.
   This may help one to figure out what kind of excess of the CR flux
   is detected and which is not.
   In particular, this may help to explain why an excess of
   cosmic ray flux is detected in direction to certain possible CR
   sources while to others it is not.
   Next, it is desirable to study arrival directions
   of CRs just above the knee with greater statistics.
   To implement this task, we plan to use the data obtained in late
   1980's with the EAS MSU Array and to employ another data set
   already obtained with the EAS--1000 Prototype Array.
   Finally, we also plan to study another interesting feature of
   the CR arrival directions revealed during the investigation.
   These are zones with `depressed' flux, i.e., zones with a
   number of EAS much less than the average.
   We think that an analysis of such `sinks' can provide some
   information useful for understanding the origin of cosmic rays in the
   energy range near the knee.


\asection{Acknowledgments}
   We thank A. A. Chilingarian, A.~D. Erlykin, and E.~Parizot for useful
   and stimulating communications.
   The whole investigation would have been impossible without our colleagues
   who obtained the experimental data.
   MZ thanks A.~Golubin, A.~Habib-ur-Rahman-Khan, V.~Lebedev, I.~Sinelobov,
   and R.~Zhukov for the nice PC, which has done the main work, and
   B.~and C.~Woodworth for the enlightening discussion.
   
   This research has made intensive use of the NASA/IPAC Extragalactic
   Database (NED), which is operated by the Jet Propulsion Laboratory,
   California Institute of Technology, under contract with the National
   Aeronautics and Space Administration, resources provided by
   SEDS~\cite{SEDS}, and especially the SIMBAD database, operated at
   CDS, Strasbourg, France.

   Only free, open source software has been used for the investigation.

\newpage
\listrefs
\bye